\documentclass[12pt]{article}

\usepackage{scicite}

\usepackage{times}
\usepackage{graphicx}
\usepackage{amsmath,amssymb}
\usepackage{url}
\usepackage{xcolor}
\usepackage{color}
\usepackage{pdfpages}
\usepackage[colorlinks=true,citecolor=blue]{hyperref}
\hypersetup{linkbordercolor=green}
\usepackage[tableposition=top]{caption}

\usepackage{tabularray}

\usepackage{arydshln}
\usepackage{comment}

\usepackage{booktabs}
\usepackage[normalem]{ulem}

\usepackage{multirow}
\usepackage{bm}     
\usepackage{enumerate}
\usepackage{subfig}
\usepackage{pict2e}

\def\hlinewd#1{%
\noalign{\ifnum0=`}\fi\hrule \@height #1 %
\futurelet\reserved@a\@xhline}

\newcommand{\ii}{{{\rm i}}}

\newenvironment{sciabstract}{%
\begin{quote} \bf}
{\end{quote}}

\usepackage{verbatim}

\title{The Sun's differential rotation is controlled by high-latitude baroclinically unstable inertial modes} 

\author{Yuto~Bekki,$^{1}$
  Robert~H.~Cameron,$^{1}$
  Laurent~Gizon$^{1,2,3,\ast}$ 
\\
\\
\normalsize{$^{1}$Max-Planck-Institut f\"ur Sonnensystemforschung,  37077 G\"ottingen, Germany}\\
\normalsize{$^{2}$Institut f\"ur Astrophysik, Georg-August-Universit\"at G\"ottingen, 37077 G\"ottingen, Germany}\\  
\normalsize{$^{3}$Center for Space Science,  New York University Abu Dhabi, Abu Dhabi, UAE}\\
\\
\normalsize{$^{\ast}$Corresponding author. Email: \href{mailto:gizon@mps.mpg.de}{gizon@mps.mpg.de} }
}

\date{}

\begin{document} 


\baselineskip24pt
\maketitle 


\begin{sciabstract}

Rapidly rotating fluids have a rotation profile which depends only on the distance from the rotation axis, in accordance with the Taylor-Proudman theorem. Although the Sun was expected to be such a body, helioseismology showed that the rotation rate in the convection zone is closer to constant on radii. It has been postulated that this deviation is due to the poles being warmer than the equator by a few degrees. Using numerical simulations, we show that the pole-to-equator temperature difference cannot exceed 7 Kelvin as a result of the back-reaction of the high-latitude baroclinically unstable inertial modes. The observed amplitudes of the modes further indicate that this maximum temperature difference is reached in the Sun. We conclude that the Sun's latitudinal differential rotation reaches its maximum allowed value.
\end{sciabstract}

\newpage

\section*{INTRODUCTION}

Global helioseismology has revealed the internal rotation profile of the Sun by analyzing the acoustic oscillations seen at the surface \cite{schou1998,howe2005}.
Throughout the convective envelope, the equator rotates faster than the poles by about 30\%.
This differential rotation  plays a crucial role in driving the solar magnetic activity via  dynamo processes \cite{charbonneau2020}.
One of the most striking features of the solar differential rotation (Fig.~\ref{fig:1}A) is a deviation from the well-known Taylor-Proudman theorem which predicts constant rotation rates on cylinders in a dynamical regime dominated by rotation \cite{Greenspan1968, Pedlosky1982}.
There are several, mutually compatible and related,  possible reasons  the theorem does not apply. 
One is that the convective velocities lead to appreciable anisotropic Reynolds stresses \cite{rudiger1989, Brun2010}. 
Another possibility is that magnetic stresses play a role \cite{hotta2018,hotta2022}. 
Yet another is that surfaces of constant density and constant pressure do not coincide in the Sun, indicating the presence of a latitudinal entropy gradient and a meridional flow \cite{kitchatinov1995,rempel2005,miesch2006}. 
It has been established with numerical simulations that, among these proposed mechanisms, the latitudinal entropy gradient is the dominant cause of the departure of the Sun's differential rotation from the Taylor-Proudman state \cite{hotta2022,miesch2006,brun2011}.
The entropy and corresponding temperature profiles
required to produce the observed differential rotation can be estimated based on the thermal wind balance approximation (Figs.~\ref{fig:1}B and \ref{fig:1}C), with the poles being warmer than the equator by about $5$~K in the middle  convection zone \cite{miesch2005}. 
Unfortunately this latitudinal temperature difference is too small to be measured by direct observation \cite{kuhn1985,woodard2003,rast2008} and our understanding of the interaction between rotation and turbulent thermal convection in the Sun is too incomplete to deduce the solar differential rotation from first principles. 
Thus, despite its crucial importance for the internal dynamo mechanism, the physical origin of the Sun's differential rotation remains an open question. 

The Sun's differential rotation gives rise to a rich spectrum of quasi-toroidal inertial modes, which have been observed on the Sun \cite{gizon2021}. 
These modes propagate in the retrograde direction in a frame rotating at the equatorial rotation rate; they include equatorial Rossby modes \cite{loeptien2018}, critical-latitude and high-latitude modes \cite{gizon2021}, and additional modes \cite{Hanson2022}.
The amplitudes of most of these modes are  consistent with stochastic excitation by turbulent convection \cite{bekki2022b,philidet2023}. 
However the high-latitude modes with low azimuthal orders $m$ are linearly unstable \cite{fournier2022,bekki2022a} and have large velocity amplitudes \cite{gizon2021}. 
We here show that these high-latitude modes play a controlling role in determining the Sun's differential rotation.

\section*{RESULTS}

\subsection*{The Sun's high-latitude modes of oscillation}

Near the poles of the Sun, a spiral pattern in the longitudinal flow has been shown to exist, predominantly with  azimuthal order $m=1$ \cite{hathaway2013}. 
This flow pattern is the surface manifestation of a  global mode of oscillation of the whole convection zone \cite{gizon2021}.  
The power spectrum of the $m=1$ longitudinal component of the surface velocity measured using SDO/HMI observations \cite{gizon2021} is shown over a range of latitudes  in Fig.~\ref{fig:1}D.  
Excess power is clearly seen above $60^\circ$ in latitude at frequency of about $-86$~nHz in the Carrington frame. 
Excess power at this frequency is also present at all other latitudes (Fig.~\ref{fig:1}E), confirming the global nature of the mode.
Following the  same procedure and data, we extended the analysis to characterize the high-latitude modes with $m=2$ and $3$.
We restricted the data to the period from 2017-2021 when the solar magnetic activity is low. 
We do this because we will compare the observed properties of the high-latitude inertial modes to those from hydrodynamic (non-magnetic) simulations.

The frequencies for the modes with $m=1$, $2$ and $3$ are $-86.3 \pm 1.6$~nHz, $-151.1 \pm 4.3$~nHz and $-224.7 \pm 2.5$~nHz, respectively.
Figures~\ref{fig:1}G, H, and I show the corresponding velocity eigenfunctions.
The $m=1$ mode has the largest amplitude among all the observed inertial modes, with a longitudinal component of velocity $v_{\phi} = 11.8 \pm 2.4$~m~s$^{-1}$.
The $m=2$ and $m=3$ high-latitude  modes have amplitudes of $3.8$~m~s$^{-1}$ and $1.9$~m~s$^{-1}$ respectively (fig.~\ref{fig_s:vobs_m123}).

Using a linear eigenvalue solver [see Supplementary Materials; \cite{bekki2022a}], we find a clear correspondence between the three observed high-latitude modes and the eigenmodes of a model of the solar convection zone including the solar differential rotation.
The properties of the high-latitude modes strongly depend on the profiles of the solar differential rotation \cite{fournier2022} and the latitudinal entropy gradient \cite{bekki2022a}. 
We find that the high-latitude modes with $m=1$, $2$ and $3$ are linearly unstable to a baroclinic instability when a latitudinal entropy gradient is applied (fig.~\ref{fig_s:linana_dispersion}). 
It is also shown that these high-latitude modes transport heat equatorward (fig.~\ref{fig_s:linana_Fe}).
The amplitudes of the modes observed on the Sun can only be understood in terms of the nonlinear evolution of the modes.

\subsection*{3D simulations of the large-scale solar dynamics}

To simulate the nonlinear mode dynamics, we use the mean-field hydrodynamic framework (see Supplementary Materials) where small-scale convection is not explicitly solved for, but the effects of the small-scale convection are included as a subgrid-scale turbulent viscosity and  momentum transport mechanism via the $\Lambda$~effect \cite{rudiger1989}. 
The use of the mean-field approach is currently the best way to closely reproduce the solar differential rotation in numerical simulations.
Here, the problem consists of the continuity, momentum and energy equations, together with an equation of state, to be solved in a three-dimensional (3D) spherical shell extending in radius from $0.65 R_\odot$ to $0.985 R_\odot$.
The perturbations of pressure, entropy and density are assumed to be small so that a linearized equation of state is used. 
To speed up the computations, we adopt the reduced speed of sound approximation \cite{rempel2005,hotta2014a}, which leads to a modified equation of continuity (eq.~\ref{eq_s:mass2}). 
We refer the reader to Supplementary Materials for an explicit description of the equations and to Materials and Methods for a summary of the numerical method.
The magnetic field is not included in this study; potential implications of the magnetic field are discussed in  Supplementary Materials.

In our simulations, non-axisymmetric inertial modes as well as realistic differential rotation and meridional circulation can be modelled (except for the near-surface shear layer).
The latitudinal entropy and temperature gradients are self-consistently generated by the radial penetration of the meridional flow into a weakly subadiabatic (stably stratified) layer at the base of the convection zone \cite{rempel2005}.
Here, the deviation from the adiabatic stratification is measured by the superadiabaticity $\delta=\nabla-\nabla_{\mathrm{ad}}$ where $\nabla=\mathrm{d}(\ln T)/\mathrm{d}(\ln p)$ is the double-logarithmic temperature gradient.
The subadiabaticity at the base of the convection zone, $\delta_{0}$, can be used to change the baroclinicity in the convection zone \cite{rempel2005}. 
Random fluctuations are added in the $\Lambda$~effect to mimic stochastic convective motions.  
All the simulations  began with uniform rotation, and were allowed to run for more than 10 simulated years after a statistically stationery state was reached.

As summarized in Table~\ref{table:1}, we carried out a total of ten 3D simulations (labelled 1 through 10) where $\delta_{0}$ was changed from $-2\times 10^{-6}$ to $-6\times 10^{-5}$ (see fig.~\ref{fig_s:delta_nonlin}). 
For the intermediate cases (cases 4, 5, and 6), a solar-like differential rotation compatible with helioseismology was found. 
For each 3D simulation, a reference 2D axisymmetric simulation was performed where the non-axisymmetric modes cannot exist, allowing us to study the role played by the non-axisymmetric modes in the 3D simulations.

Figure~\ref{fig:2} shows the temporal evolution of the large-scale flows in the convection zone. 
Since the initial conditions correspond to solid body rotation, there is at first no differential rotation, no latitudinal entropy variation (baroclinicity), and no power in the high-latitude modes.
As time progresses, the axisymmetric mean flows and the mean baroclinicity are established, and the modes grow in amplitude at high latitudes (figs.~\ref{fig_s:final_vphi}--\ref{fig_s:final_s1}). 
The modes initially have little influence on the evolution of the mean state. 
This can be inferred from the nearly identical evolution in the 2D and 3D simulations. 
As the mode amplitudes grow, we see that the temporal behaviour of the high-latitude modes is non-linearly coupled with the mean background state. 
This is seen by the large differences between the results of the 2D axisymmetric and 3D non-axisymmetric simulations.

For the 2D axisymmetric simulations,  Fig.~\ref{fig:2} shows that the final value of $\Delta_{\theta}\overline{s}_{1}=\overline{s}_{1,\mathrm{eq}}-\overline{s}_{1,\mathrm{pole}}$
decreases from case 1 to 10 corresponding to an increase in the baroclinicity.
Note that the bar on top of quantities indicates a longitudinal average and $\Delta_{\theta}$ denotes the difference equator minus poles throughout this paper.
Related to the baroclinicity increase, we find a monotonic increase of  the latitudinal differential rotation 
$\Delta_{\theta}\Omega = \Omega_{\mathrm{eq}}-\Omega_{\mathrm{pole}}$
from case 1 to 10.
The situation is very different in the 3D simulations where the high-latitude modes are allowed (figs.~\ref{fig_s:Feq}--\ref{fig_s:final_MC}).
In all cases, $\Delta_{\theta}\Omega$ and $\Delta_{\theta}\overline{s}_{1}$ are reduced in amplitude compared to their 2D counterparts at the end of the simulations.
In cases 1--6, the reduction is modest.
However, in cases 7--10, we find that both $\Delta_{\theta}\Omega$ and $|\Delta_{\theta}\overline{s}_{1}|$ undergo a substantial reduction before reaching the statistically-stationary states at levels substantially smaller than those in the 2D cases. 
The strong reduction in $|\Delta_{\theta}\overline{s}_{1}|$ in these cases is caused by the equatorward heat transport by the high-latitude modes which have large velocity amplitudes, in excess of $35$~m~s$^{-1}$ (see fig.~\ref{fig_s:Feq}).
The change in $\Delta_{\theta}\overline{s}_{1}$ produces a change in the meridional circulation (see Supplementary Materials).
The strong reduction in $\Delta_{\theta}\Omega$ is then a consequence of the change in poleward angular momentum transport by the meridional circulation (see fig.~\ref{fig_s:instability_case6}). {The direct transport of angular momentum by the modes is not the dominant effect, contrary to the shallow-water case studied in Ref.~\cite{dikpati2018}.} 

It is found that, when the high-latitude modes are strong, the poleward meridional flow at the surface extends towards higher latitudes, forming a clear single-cell pattern in each hemisphere (fig.~\ref{fig_s:final_MC}). 
In all cases, the amplitude of the flow at the surface is on the order of $15$~m~s$^{-1}$. 
The meridional flow is thus consistent with the results from local helioseismology \cite{gizon2020s}.

\section*{DISCUSSION}

\subsection*{Constraints on the latitudinal temperature difference}

The amplitudes of the observed $m=1$ high-latitude mode and the observed latitudinal differential rotation are reproduced by our nonlinear simulation cases 4, 5, and 6.
Figures~\ref{fig:3}A--C show the differential rotation $\Omega$, the mean entropy perturbation $\overline{s}_{1}$, and the mean temperature perturbation $\overline{T}_{1}$ from our reference case 5.
Figures~\ref{fig:3}D and E show the $m=1$ power spectra of $v_{\phi}$ at the surface.
Although the mode power is mostly concentrated at high latitudes (above $60^{\circ}$), the power can be seen at all latitudes at the same frequency once it is normalized.
This is consistent with the solar observations (Fig.~\ref{fig:1}D and E).
Furthermore, Fig.~\ref{fig:3}F shows that a ridge of the high-latitude velocity power closely follows the observed propagation frequencies of the high-latitude modes (Fig.~\ref{fig:1}F).
We use the singular-value-decomposition method \cite{bekki2022b} to extract the spatial eigenfunctions of the low-$m$ high-latitude modes (figs.~\ref{fig_s:nonlin_eigfunc} and \ref{fig_s:eigenfunc_case4}).
The surface eigenfunctions of the $m=1$, $2$, and $3$ modes are presented in Figs.~\ref{fig:3}G--I. 
The observed spiral pattern of the high-latitude modes can be nicely reproduced (Figs.~\ref{fig:1}G--I).
The amplitude of the $m=1$ mode at the surface match the observed value very well (Table~\ref{table:1}).

The observed amplitudes of the high-latitude modes place clear limits on $\Delta_{\theta}\overline{T}_{1}$. 
We see in Figs.~\ref{fig:2}A and \ref{fig:4}B that the amplitude of the $m=1$ mode is too high in case~7 and too low in case~3, while cases~4--6 give amplitudes of order $10$--$15$~m~s$^{-1}$. 
A similar constraint is obtained by comparing the observed and simulated amplitudes of the mode with $m=2$.
The mode with $m=3$ does not provide a tight constraint.
From Table~\ref{table:1} and Fig.~\ref{fig:4}, we deduce $-7.0$~K $< \Delta_{\theta}\overline{T}_{1} <-6.8$~K in the middle convection zone. 
In case 5, which best matches the solar observations, the entropy difference between the cooler equator and the warmer poles in the middle convection zone is $\Delta_{\theta}\overline{s}_{1} = -981$ erg~g$^{-1}$~K$^{-1}$ and the temperature difference is $\Delta_{\theta}\overline{T}_{1} = -6.8$~K. 
This temperature difference in the Sun is much too small to be measured by more direct means.

\subsection*{The role of the high-latitude inertial modes in determining the Sun's differential rotation}

The mean latitudinal temperature gradient plays a key role in determining the Sun's differential rotation by driving the meridional circulation which redistributes the angular momentum.
The differential rotation profile in a statistically stationary state is determined by a balance between the Coriolis force and the latitudinal pressure gradient force, which leads to the equation for thermal wind balance (eq.~\ref{eq_s:dsdq}). 
The inferred value of $\Delta_{\theta}\overline{T}_{1} = -6.8$~K is consistent with that required for the observed differential rotation to be in thermal wind balance \cite{miesch2005}.

The latitudinal temperature gradient also drives the non-axisymmetric high-latitude inertial modes which are baroclinically unstable at low $m$. 
In the presence of a sufficiently strong latitudinal temperature difference, the low-$m$ modes, especially the $m=1$ mode, grow in amplitude and begin transporting heat towards the equator (see figs.~\ref{fig_s:linana_Fe}, ~\ref{fig_s:Feq}, and \ref{fig_s:RSFehk_case4}). 
This reduces the temperature of the poles relative to the equator so that the modes are less strongly driven. 
The modes saturate by reducing the background baroclinicity.
The value of $\Delta_{\theta}\overline{T}_{1}$ in a statistically stationary state reflects this balance and thus is determined by the nonlinear heat transport by the high-latitude inertial modes.

Unexpectedly, our study reveals that the latitudinal temperature difference in the Sun ($-6.8$~K) is near its maximum possible (absolute) value, see 3D case in Fig.~\ref{fig:4}A.
In our 2D axisymmetric simulations, it is possible to achieve a higher baroclinicity and latitudinal temperature difference
by increasing $|\delta_0|$, see 2D case in Fig.~\ref{fig:4}A. 
This is not possible in the 3D simulations where the non-axisymmetric high-latitude modes are present.   
With increasing $|\delta_0|$, the high-latitude inertial modes grow in amplitude but the temperature gradient does not increase because of enhanced equatorward heat transport by the modes.
The modes change the baroclinicity and thus affect the meridional flow, resulting in poleward angular momentum transport (fig.~\ref{fig_s:instability_case6}).
This indirect effect is larger than the direct transport of angular momentum  by the modes, which is equatorward (fig.~\ref{fig_s:RSFehk_case4}).

The latitudinal differential rotation thus has a maximum achievable value 
when the non-axisymmetric inertial  modes are included (the 3D case in Fig.~\ref{fig:4}C).
Furthermore we find that the maximum differential rotation in our nonlinear simulations is $\Delta_\theta \Omega / 2\pi =134.7$~nHz  (case 5), which is very close to the observed value of $130.8 \pm 4.5$~nHz (see Fig.~\ref{fig:4}C).
We conclude that the high-latitude modes explain the Sun's differential rotation.
Whether the same mechanism plays a significant role in explaining the large latitudinal differential rotation observed in faster rotating stars \cite{benomar2018} is an open question.

\section*{MATERIALS AND METHODS}

\subsection*{Nonlinear 3D dynamical model}

In order to study the baroclinic excitation and the nonlinear saturation of the high-latitude modes, we carry out a set of 3D hydrodynamic simulations of the large-scale flows in the Sun.
We  use a mean-field approach where the time-averaged convective angular momentum transport is parameterized by the $\Lambda$ effect (see Supplementary Materials).
This enables us to study the high-latitude modes in the context of a solar-like differential rotation profile.
 
We solve the mean-field equations of motion and energy, together with a modified continuity equation. 
To speed-up the computations, we follow the method suggested by Ref.~\cite{rempel2005}.
This leads to a reduction of the effective sound speed by a factor of $\xi = 150$ and a relaxation of the Courant-Friedrichs-Lewy (CFL) condition.
We use the same background stratification as in Ref.~\cite{rempel2005}.

We include a weakly-subadiabatic overshooting layer. The meridional circulation penetrates into this layer driving a latitudinal entropy gradient (baroclinicity) in the convection zone \cite{rempel2005,brun2011}.
The subadiabaticity $\delta_0$ in the overshooting layer is a variable parameter. 

For numerically solving the governing equations, we use the hydrodynamic solver of the mean-field MHD dynamo code described in Ref.~\cite{bekki2023}.
The numerical scheme consists of a $4$th-order centered-difference method for space and a $4$-step Runge-Kutta scheme for the time integration \cite{voegler2005}.
The numerical domain extends from $0.65R_{\odot}$ up to $0.985R_{\odot}$.
For simplicity, the near-surface shear layer is not included.
At both radial boundaries, an impenetrable and stress-free boundary condition is assumed.
A Yin-Yang grid is used to avoid the singularities of the spherical coordinate \cite{bekki2022b,kageyama2004}.
The grid resolution used in this study is $N_r\times N_\theta \times N_\phi = 72\times 188\times 382$.
For the initial conditions, we set all the variables (velocity, density, and entropy fluctuations) to zero.

\vspace{1cm}
\section*{Supplementary Materials}

\begin{itemize}
\item[-] Supplementary Text
\item[-] Figs.~\ref{fig_s:vobs_m123}--\ref{fig_s:fig4_lambda}
\item[-] Table~\ref{table_s:lambda}
\end{itemize}

\vspace{1cm}
\bibliography{ref}

\begin{thebibliography}{10}

\bibitem{schou1998}
J.~{Schou}, H.~M. {Antia}, S.~{Basu}, R.~S. {Bogart}, R.~I. {Bush}, S.~M.
  {Chitre}, J.~{Christensen-Dalsgaard}, M.~P. {Di Mauro}, W.~A. {Dziembowski},
  A.~{Eff-Darwich}, D.~O. {Gough}, D.~A. {Haber}, J.~T. {Hoeksema}, R.~{Howe},
  S.~G. {Korzennik}, A.~G. {Kosovichev}, R.~M. {Larsen}, F.~P. {Pijpers}, P.~H.
  {Scherrer}, T.~{Sekii}, T.~D. {Tarbell}, A.~M. {Title}, M.~J. {Thompson},
  J.~{Toomre}, {Helioseismic Studies of Differential Rotation in the Solar
  Envelope by the Solar Oscillations Investigation Using the Michelson Doppler
  Imager}.
\newblock {\it \apj\/} {\bf 505}, 390-417 (1998).

\bibitem{howe2005}
R.~{Howe}, J.~{Christensen-Dalsgaard}, F.~{Hill}, R.~{Komm}, J.~{Schou}, M.~J.
  {Thompson}, {Solar Convection-Zone Dynamics, 1995-2004}.
\newblock {\it \apj\/} {\bf 634}, 1405-1415 (2005).

\bibitem{charbonneau2020}
P.~{Charbonneau}, {Dynamo models of the solar cycle}.
\newblock {\it Living Reviews in Solar Physics\/} {\bf 17}, 4 (2020).
  \href{https://doi.org/10.1007/s41116-020-00025-6}{https://doi.org/10.1007/s41116-020-00025-6}.

\bibitem{Greenspan1968}
H.~Greenspan, {\it The Theory of Rotating Fluids\/}, Cambridge Monographs on
  Mechanics (Cambridge University Press, 1968).

\bibitem{Pedlosky1982}
J.~{Pedlosky}, {\it {Geophysical Fluid Dynamics}\/} (Springer-Verlag New York,
  1982).

\bibitem{rudiger1989}
G.~{R\"udiger}, {\it {Differential Rotation and Stellar Convection}\/} (Berlin:
  Akademie Verlag, 1989).

\bibitem{Brun2010}
A.~S. {Brun}, H.~M. {Antia}, S.~M. {Chitre}, {Is the solar convection zone in
  strict thermal wind balance?}
\newblock {\it \aap\/} {\bf 510}, A33 (2010).

\bibitem{hotta2018}
H.~{Hotta}, {Breaking Taylor-Proudman Balance by Magnetic Fields in Stellar
  Convection Zones}.
\newblock {\it \apjl\/} {\bf 860}, L24 (2018).

\bibitem{hotta2022}
H.~{Hotta}, K.~{Kusano}, R.~{Shimada}, {Generation of Solar-like Differential
  Rotation}.
\newblock {\it \apj\/} {\bf 933}, 199 (2022).

\bibitem{kitchatinov1995}
L.~L. {Kitchatinov}, G.~{R\"udiger}, {Differential rotation in solar-type
  stars: revisiting the Taylor-number puzzle.}
\newblock {\it \aap\/} {\bf 299}, 446 (1995).

\bibitem{rempel2005}
M.~{Rempel}, {Solar Differential Rotation and Meridional Flow: The Role of a
  Subadiabatic Tachocline for the Taylor-Proudman Balance}.
\newblock {\it \apj\/} {\bf 622}, 1320-1332 (2005).

\bibitem{miesch2006}
M.~S. {Miesch}, A.~S. {Brun}, J.~{Toomre}, {Solar Differential Rotation
  Influenced by Latitudinal Entropy Variations in the Tachocline}.
\newblock {\it \apj\/} {\bf 641}, 618-625 (2006).

\bibitem{brun2011}
A.~S. {Brun}, M.~S. {Miesch}, J.~{Toomre}, {Modeling the Dynamical Coupling of
  Solar Convection with the Radiative Interior}.
\newblock {\it \apj\/} {\bf 742}, 79 (2011).

\bibitem{miesch2005}
M.~S. {Miesch}, {Large-Scale Dynamics of the Convection Zone and Tachocline}.
\newblock {\it Living Reviews in Solar Physics\/} {\bf 2}, 1 (2005).
  \href{https://doi.org/10.12942/lrsp-2005-1}{https://doi.org/10.12942/lrsp-2005-1}.

\bibitem{kuhn1985}
J.~R. {Kuhn}, K.~G. {Libbrecht}, R.~H. {Dicke}, {Observations of a Solar
  Latitude-dependent Limb Brightness Variation}.
\newblock {\it \apj\/} {\bf 290}, 758 (1985).

\bibitem{woodard2003}
M.~F. {Woodard}, K.~G. {Libbrecht}, {Spatial and temporal variations in the
  solar brightness}.
\newblock {\it \solphys\/} {\bf 212}, 51-64 (2003).

\bibitem{rast2008}
M.~P. {Rast}, A.~{Ortiz}, R.~W. {Meisner}, {Latitudinal Variation of the Solar
  Photospheric Intensity}.
\newblock {\it \apj\/} {\bf 673}, 1209-1217 (2008).

\bibitem{gizon2021}
L.~{Gizon}, R.~H. {Cameron}, Y.~{Bekki}, A.~C. {Birch}, R.~S. {Bogart}, A.~S.
  {Brun}, C.~{Damiani}, D.~{Fournier}, L.~{Hyest}, K.~{Jain}, B.~{Lekshmi},
  Z.-C. {Liang}, B.~{Proxauf}, {Solar inertial modes: Observations,
  identification, and diagnostic promise}.
\newblock {\it \aap\/} {\bf 652}, L6 (2021).

\bibitem{loeptien2018}
B.~{L{\"o}ptien}, L.~{Gizon}, A.~C. {Birch}, J.~{Schou}, B.~{Proxauf}, T.~L.
  {Duvall Jr.}, R.~S. {Bogart}, U.~R. {Christensen}, {Global-scale equatorial
  Rossby waves as an essential component of solar internal dynamics}.
\newblock {\it Nature Astronomy\/} {\bf 2}, 568-573 (2018).

\bibitem{Hanson2022}
C.~S. {Hanson}, S.~{Hanasoge}, K.~R. {Sreenivasan}, {Discovery of
  high-frequency retrograde vorticity waves in the Sun}.
\newblock {\it Nature Astronomy\/} {\bf 6}, 708-714 (2022).

\bibitem{bekki2022b}
Y.~{Bekki}, R.~H. {Cameron}, L.~{Gizon}, {Theory of solar oscillations in the
  inertial frequency range: Amplitudes of equatorial modes from a nonlinear
  rotating convection simulation}.
\newblock {\it \aap\/} {\bf 666}, A135 (2022).

\bibitem{philidet2023}
J.~{Philidet}, L.~{Gizon}, {Interaction of solar inertial modes with turbulent
  convection. A 2D model for the excitation of linearly stable modes}.
\newblock {\it \aap\/} {\bf 673}, A124 (2023).

\bibitem{fournier2022}
D.~{Fournier}, L.~{Gizon}, L.~{Hyest}, {Viscous inertial modes on a
  differentially rotating sphere: Comparison with solar observations}.
\newblock {\it \aap\/} {\bf 664}, A6 (2022).

\bibitem{bekki2022a}
Y.~{Bekki}, R.~H. {Cameron}, L.~{Gizon}, {Theory of solar oscillations in the
  inertial frequency range: Linear modes of the convection zone}.
\newblock {\it \aap\/} {\bf 662}, A16 (2022).

\bibitem{hathaway2013}
D.~H. {Hathaway}, L.~{Upton}, O.~{Colegrove}, {Giant Convection Cells Found on
  the Sun}.
\newblock {\it Science\/} {\bf 342}, 1217-1219 (2013).

\bibitem{hotta2014a}
H.~{Hotta}, M.~{Rempel}, T.~{Yokoyama}, {High-resolution Calculations of the
  Solar Global Convection with the Reduced Speed of Sound Technique. I. The
  Structure of the Convection and the Magnetic Field without the Rotation}.
\newblock {\it \apj\/} {\bf 786}, 24 (2014).

\bibitem{dikpati2018}
M.~{Dikpati}, S.~W. {McIntosh}, G.~{Bothun}, P.~S. {Cally}, S.~S. {Ghosh},
  P.~A. {Gilman}, O.~M. {Umurhan}, {Role of Interaction between Magnetic Rossby
  Waves and Tachocline Differential Rotation in Producing Solar Seasons}.
\newblock {\it \apj\/} {\bf 853}, 144 (2018).

\bibitem{gizon2020s}
L.~{Gizon}, R.~H. {Cameron}, M.~{Pourabdian}, Z.-C. {Liang}, D.~{Fournier},
  A.~C. {Birch}, C.~S. {Hanson}, {Meridional flow in the Sun{\textquoteright}s
  convection zone is a single cell in each hemisphere}.
\newblock {\it Science\/} {\bf 368}, 1469-1472 (2020).

\bibitem{benomar2018}
O.~{Benomar}, M.~{Bazot}, M.~B. {Nielsen}, L.~{Gizon}, T.~{Sekii}, M.~{Takata},
  H.~{Hotta}, S.~{Hanasoge}, K.~R. {Sreenivasan}, J.~{Christensen-Dalsgaard},
  {Asteroseismic detection of latitudinal differential rotation in 13 Sun-like
  stars}.
\newblock {\it Science\/} {\bf 361}, 1231-1234 (2018).

\bibitem{bekki2023}
Y.~{Bekki}, R.~H. {Cameron}, {Three-dimensional non-kinematic simulation of the
  post-emergence evolution of bipolar magnetic regions and the Babcock-Leighton
  dynamo of the Sun}.
\newblock {\it \aap\/} {\bf 670}, A101 (2023).

\bibitem{voegler2005}
A.~{V{\"o}gler}, S.~{Shelyag}, M.~{Sch{\"u}ssler}, F.~{Cattaneo}, T.~{Emonet},
  T.~{Linde}, {Simulations of magneto-convection in the solar photosphere.
  Equations, methods, and results of the MURaM code}.
\newblock {\it \aap\/} {\bf 429}, 335-351 (2005).

\bibitem{kageyama2004}
A.~Kageyama, T.~Sato, {Yin-Yang grid: An overset grid in spherical geometry}.
\newblock {\it Geochemistry, Geophysics, Geosystems\/} {\bf 5} (2004).

\bibitem{hotta2023}
H.~{Hotta}, Y.~{Bekki}, L.~{Gizon}, Q.~{Noraz}, M.~{Rast}, {Dynamics of
  Large-Scale Solar Flows}.
\newblock {\it \ssr\/} {\bf 219}, 77 (2023).

\bibitem{kuker2001}
M.~{K{\"u}ker}, M.~{Stix}, {Differential rotation of the present and the
  pre-main-sequence Sun}.
\newblock {\it \aap\/} {\bf 366}, 668-675 (2001).

\bibitem{miesch2011}
M.~S. {Miesch}, B.~W. {Hindman}, {Gyroscopic Pumping in the Solar Near-surface
  Shear Layer}.
\newblock {\it \apj\/} {\bf 743}, 79 (2011).

\bibitem{featherstone2015}
N.~A. {Featherstone}, M.~S. {Miesch}, {Meridional Circulation in Solar and
  Stellar Convection Zones}.
\newblock {\it \apj\/} {\bf 804}, 67 (2015).

\bibitem{proxauf2020}
B.~{Proxauf}, L.~{Gizon}, B.~{L{\"o}ptien}, J.~{Schou}, A.~C. {Birch}, R.~S.
  {Bogart}, {Exploring the latitude and depth dependence of solar Rossby waves
  using ring-diagram analysis}.
\newblock {\it \aap\/} {\bf 634}, A44 (2020).

\bibitem{bekki2017a}
Y.~{Bekki}, T.~{Yokoyama}, {Double-cell-type Solar Meridional Circulation Based
  on a Mean-field Hydrodynamic Model}.
\newblock {\it \apj\/} {\bf 835}, 9 (2017).

\bibitem{gilman2015}
P.~A. {Gilman}, {Effect of Toroidal Fields On Baroclinic Instability in the
  Solar Tachocline}.
\newblock {\it \apj\/} {\bf 801}, 22 (2015).

\bibitem{gilman2017}
P.~A. {Gilman}, {Baroclinic Instability in the Solar Tachocline for Continuous
  Vertical Profiles of Rotation, Effective Gravity, and Toroidal Field}.
\newblock {\it \apj\/} {\bf 842}, 130 (2017).

\end{thebibliography}
\bibliographystyle{ScienceAdvances}

\clearpage
\newpage
\section*{Acknowledgments}
\begin{figure}
\begin{center}
\vspace{-1.5cm}
\includegraphics[scale=0.6]{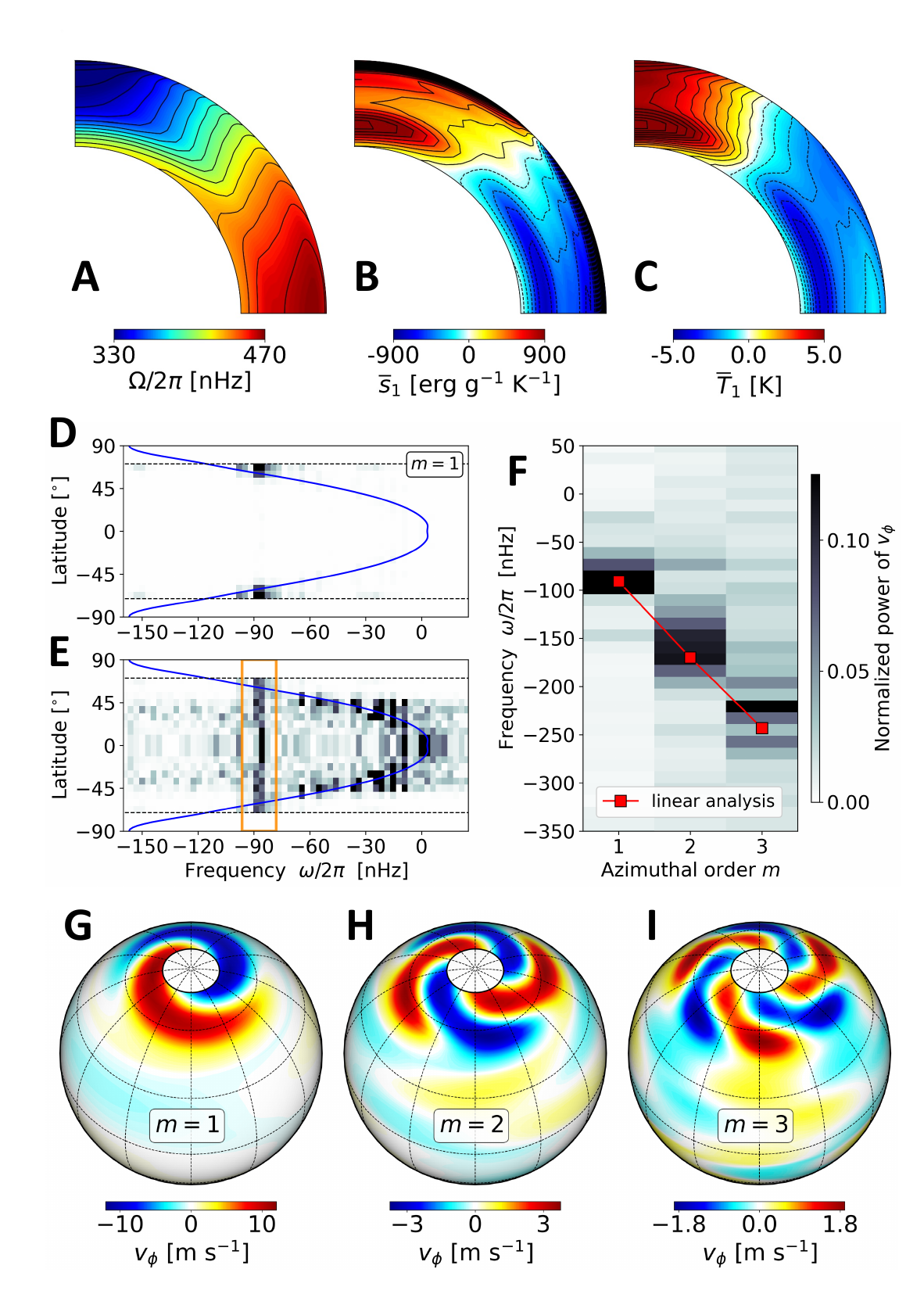}
\caption{{\bf Observations.}
({\bf A}) Solar internal rotation rate,  $\Omega(r,\theta)$, obtained by global helioseismology \cite{howe2005} using the GONG data from April 2010 to February 2021 (data courtesy of R.~Howe). 
Corresponding entropy perturbation ({\bf B}) and temperature perturbation ({\bf C}) 
required to sustain the solar differential rotation via thermal wind balance (see eq.~S4).
({\bf D}) Power spectrum of $v_{\phi}$ for the $m=1$ mode in the Carrington frame.
The blue curve shows the latitudinal differential rotation at $r=0.985R_{\odot}$.
({\bf E}) The same power spectrum but normalized at each latitude by its average over the frequency range 
between the orange bars \cite{gizon2021}.
({\bf F}) Observed dispersion relation of the high-latitude modes from the power spectra of $v_\phi$.
The theoretical dispersion relation from the linearized system is overplotted in red \cite{bekki2022a}.
({\bf G}, {\bf H}, and {\bf I}) Observed near-surface longitudinal velocity $v_{\phi}$ for the high-latitude inertial modes with $m=1$, $2$, and $3$ from ring-diagram helioseismology \cite{gizon2021}.
\label{fig:1}
  }
\end{center}
\end{figure}

\begin{figure}
\begin{center}
\includegraphics[scale=0.46]{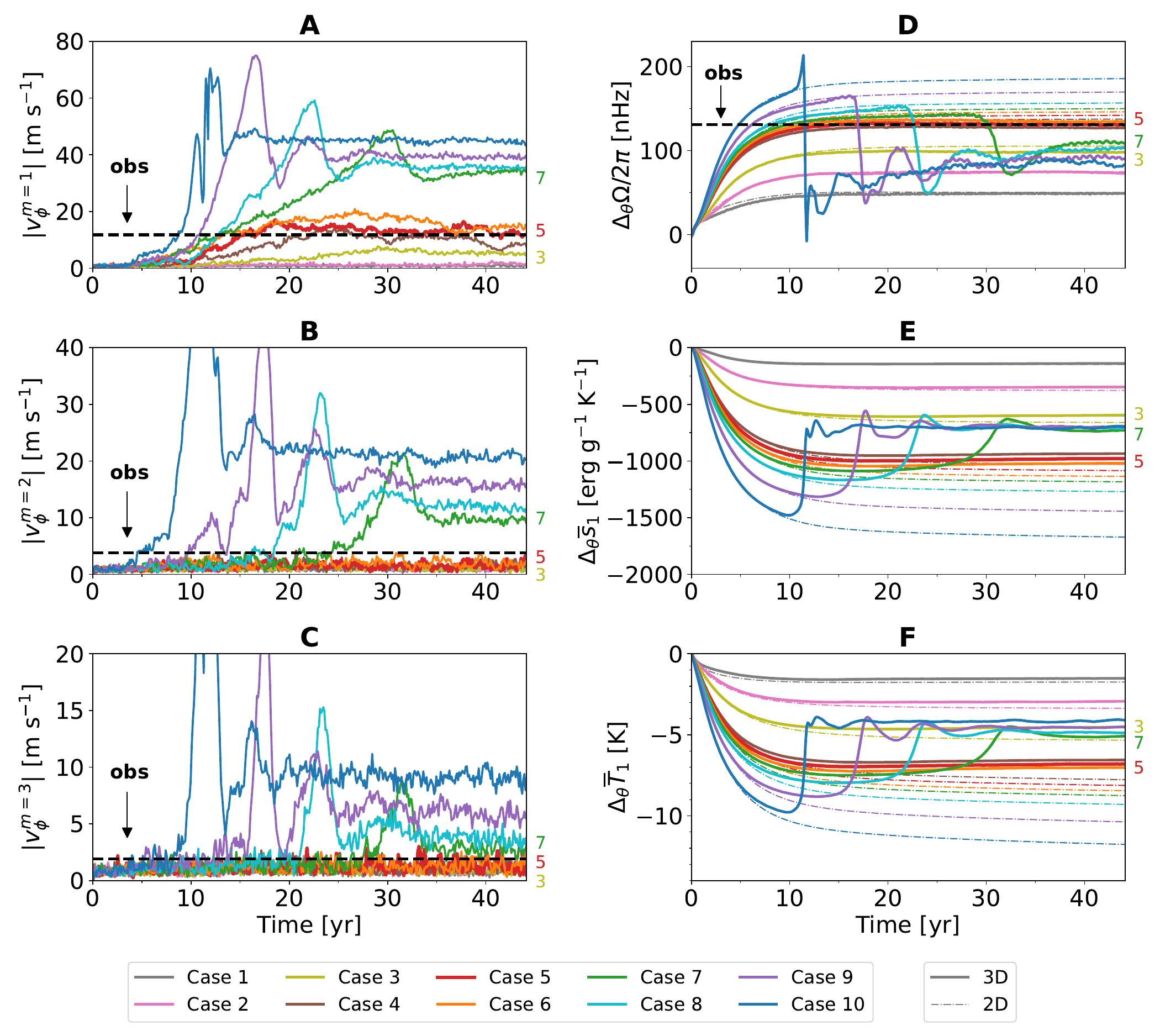}
\caption{{\bf Temporal evolution of the nonlinear simulations.} 
({\bf A}) Maximum value of $|v_{\phi}|$ for the $m=1$ high-latitude mode as a function of time.
The horizontal black dashed line shows the observed (obs) value.
The different colors refer to the different cases.
({\bf B} and {\bf C}) Same as panel~A for the $m=2$ and $m=3$ modes.
({\bf D}) Latitudinal differential rotation $\Delta_{\theta}\Omega=\Omega_{\mathrm{eq}}-\Omega_{\mathrm{pole}}$ at $r=0.85R_{\odot}$.
The solid and dashed curves denote the results from the 3D full-spherical simulations (where the high-latitude modes are present) and from the 2D axisymmetric simulations (where the high-latitude modes cannot exist), respectively. 
({\bf E}) Pole-to-equator entropy difference $\Delta_{\theta}\overline{s}_{1}=\overline{s}_{1,\mathrm{eq}}-\overline{s}_{1,\mathrm{pole}}$. 
({\bf F}) Corresponding latitudinal difference in temperature between the poles and equator $\Delta_{\theta}\overline{T}_{1}=\overline{T}_{1,\mathrm{eq}}-\overline{T}_{1,\mathrm{pole}}$.
}
\label{fig:2}
\end{center}
\end{figure}

\begin{figure}
\begin{center}
\vspace{-1.5cm}
\includegraphics[scale=0.6]{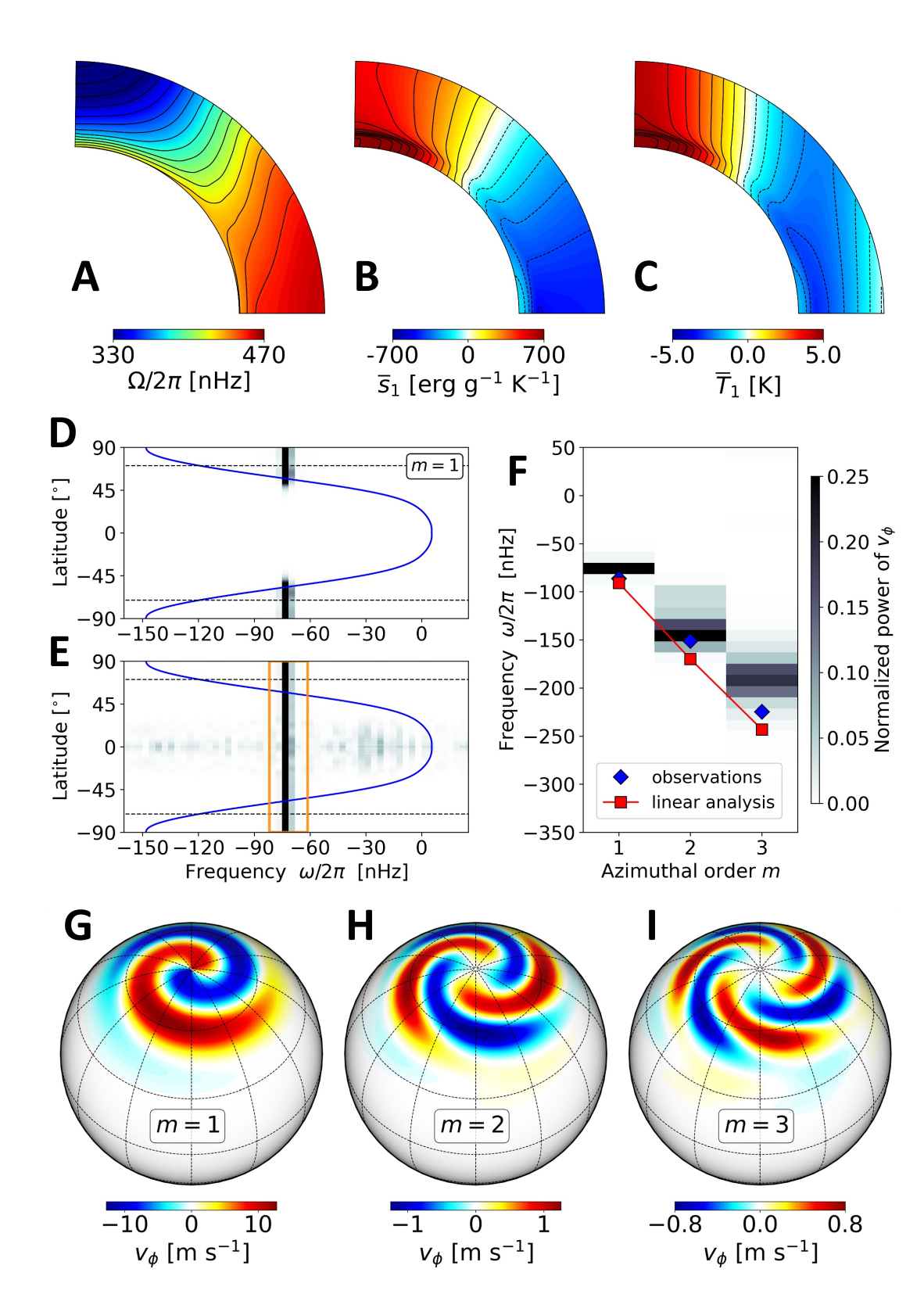}
\caption{
{\bf Results from our nonlinear simulation model (case 5).}
This figure shows the properties of the statistically stationary state of the simulated case 5. 
The figure uses the same format as used for the observations in Fig.~\ref{fig:1}. 
({\bf A}) Internal rotation rate $\Omega$, ({\bf B}) longitudinally-averaged entropy perturbation $\overline{s}_{1}$, and ({\bf C}) longitudinally-averaged temperature perturbation $\overline{T}_{1}$ produced in the nonlinear simulation.
({\bf D}) Power spectrum of $v_{\phi}$ corresponding to the high-latitude $m=1$ mode in the simulation at $r=0.985R_{\odot}$. 
The blue curve shows the latitudinal differential rotation at $r=0.985 R_\odot$. 
({\bf E}) The power spectrum normalized at each value so that the average between the orange bars is the same. 
({\bf F}) Dispersion relationship from the nonlinear simulation based on the power spectra of $v_\phi$. 
The red curve is the dispersion relation from the linearized system and the blue diamonds show the observed frequencies of the high-latitude modes \cite{gizon2021}.
({\bf G}, {\bf H}, and {\bf I}) The longitudinal velocity $v_\phi$ corresponding to the high-latitude $m=1$, $2$ and $3$ modes extracted from the upper boundary of the simulations. 
}
\label{fig:3}
\end{center}
\end{figure}

\noindent 
We are grateful to Damien Fournier, Zhi-Chao Liang, and Bastian Proxauf for useful discussions and help with the analysis of the observations.
{\bf Funding:}
We acknowledge support from ERC Synergy Grant WHOLE SUN 810218. L.G. acknowledges support from Deutsche Forschungsgemeinschaft (DFG) grant SFB 1456/432680300 Mathematics of Experiment (project C04) and NYUAD Institute grant G1502.
The computations were performed on a supercomputer of the Max Planck Society in Garching bei~M\"unchen.
{\bf Author contributions:}
All three authors designed research and wrote the paper. 
Y.B. developed the numerical codes, carried out the simulations, and analyzed the data. 
{\bf Competing interests:}
The authors declare no competing interests.
{\bf Data and materials availability:}
All data needed to evaluate the conclusions in the paper are present in the paper and/or the Supplementary Materials.
The simulation data necessary to reproduce the Figs.~\ref{fig:1}--\ref{fig:3} are available from the Edmond repository of the Max Planck Society at \url{https://doi.org/10.17617/3.SSRYIM}.
The data for reproducing Fig.~\ref{fig:4} are included in the article and in the supplementary materials.
\clearpage

\newpage
\clearpage



\begin{figure}
\begin{center}
\includegraphics[scale=0.62]{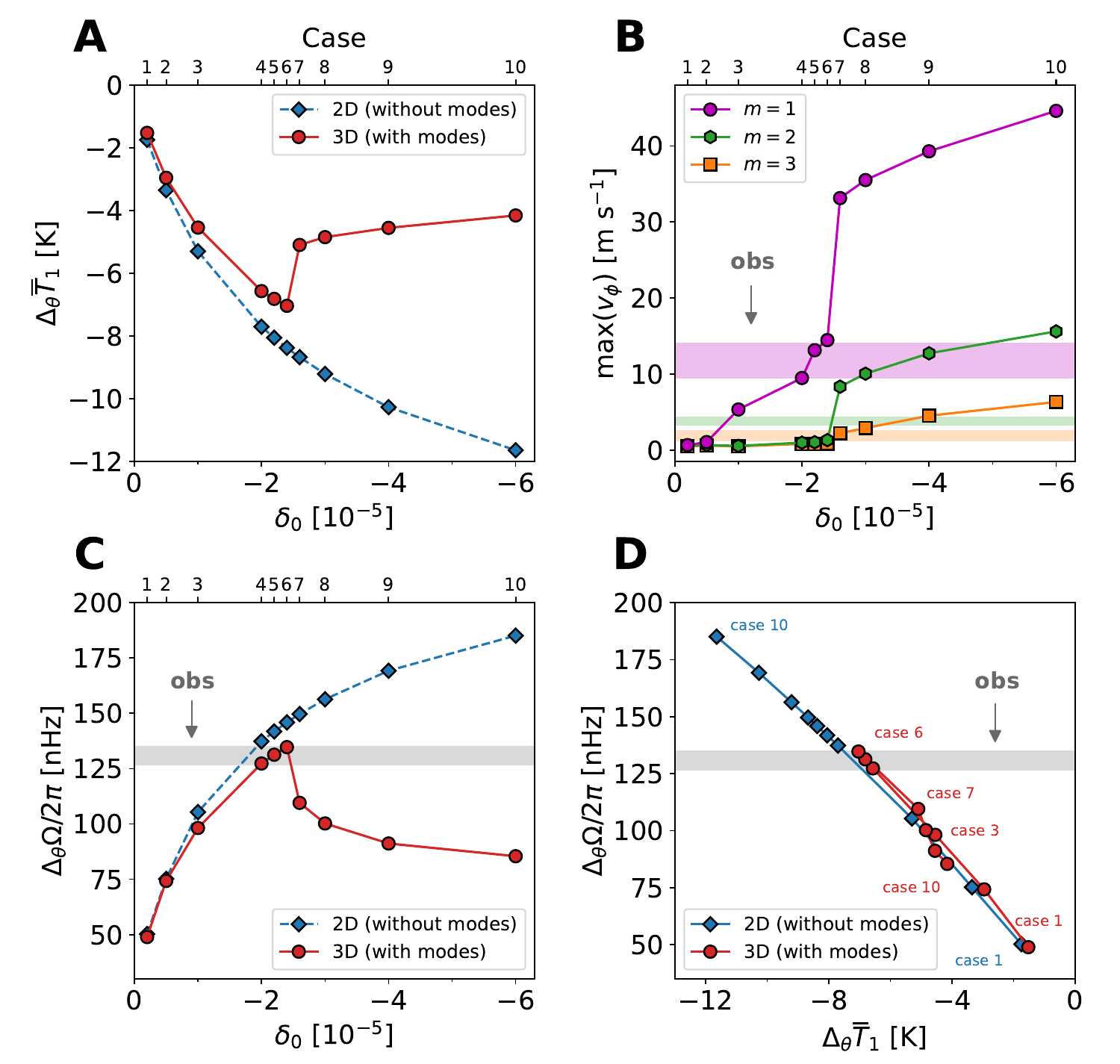}
\caption{{\bf Results of the nonlinear simulations in statistically-stationary states.}
({\bf A}) The latitudinal temperature difference $\Delta_{\theta}\overline{T}_{1}$ in the middle convection zone as a function of the superadiabaticity at the base of the convection zone $\delta_{0}$.
Blue diamonds and red circles denote the results from 2D axisymmetric simulations (without modes) and from 3D simulations (with modes), respectively.
({\bf B}) The longitudinal velocity amplitudes of the $m=1$, $2$, and $3$ modes at the surface as functions of $\delta_{0}$.
The purple, green, and orange shaded areas show the observed velocity amplitudes of the $m=1$, $2$, and $3$ modes, respectively.
({\bf C}) The latitudinal differential rotation $\Delta_{\theta}\Omega$ in the middle convection zone as a function of $\delta_{0}$.
The gray shaded area shows the observed value.
({\bf D}) Relationship between $\Delta_{\theta}\overline{T}_{1}$ and $\Delta_{\theta}\Omega$.
}
\label{fig:4}
\end{center}
\end{figure}

\newpage
\clearpage

\renewcommand{\arraystretch}{1.5}
\begin{table}[]
 \begin{center} 
\caption{
{\bf Summary of the nonlinear simulations.}
The subadiabaticity at the base of the convection zone, $\delta_{0}$, is the only free parameter in our model which controls the baroclinicity in the convection zone.
In all cases, the amplitude of the $\Lambda$~effect is fixed to $\Lambda_{0}=0.85$; see Supplementary Materials.
All the remaining quantities are results from the simulations and the values given are for the end of the simulations where statistically-stationary states have been achieved.
The quantity $\Delta_{\theta}{\Omega}={\Omega}_{\mathrm{eq}}-{\Omega}_{\mathrm{pole}}$ is the difference in the rotation rate between the poles and the equator at the middle convection zone $r=0.85R_{\odot}$.
The quantity $\Delta_{\theta}\overline{s}_{1}=\overline{s}_{1,\mathrm{eq}}-\overline{s}_{1,\mathrm{pole}}$ is the latitudinal entropy difference between the poles and the equator at the middle convection zone.
The quantity $\Delta_{\theta}\overline{T}_{1}=\overline{T}_{1,\mathrm{eq}}-\overline{T}_{1,\mathrm{pole}}$ represents the latitudinal temperature difference between the poles and the equator at the middle convection zone.
Negative signs of $\Delta_{\theta} \overline{s}_{1}$ and $\Delta_{\theta} \overline{T}_{1}$ denote that the poles are warmer than the equator.
The values in parentheses show the results from the two-dimensional (2D) axisymmetric simulations where the non-axisymmetric modes are excluded.
The quantity max($v_{\phi})$ represents the maximum longitudinal velocity amplitudes of the high-latitude modes and reported separately for $m=1$, $2$, and $3$.
The observed value for $\Delta_{\theta}\overline{\Omega}$ is obtained by global helioseismology. 
The observed value max($v_{\phi})$ is obtained from ring-diagram local helioseismology with $5^\circ$ tiles \cite{gizon2021}.
}
\scriptsize
\begin{tabular}{ccccccccccc} 
\hline
\hline
\multirow{2}{*}{Case}  & \multirow{2}{*}{$\delta_{0}$} & $\Delta_{\theta}{\Omega}/2\pi$ [nHz] 
    & $\Delta_{\theta} \overline{s}_{1}$ [erg g$^{-1}$ K$^{-1}$] 
    & $\Delta_{\theta} \overline{T}_{1}$ [K] & \multicolumn{4}{c}{max($v_{\phi}$)  [m s$^{-1}$]} \\
\cdashline{6-11}[1pt/1pt]
&& 3D (2D) & 3D (2D) & 3D (2D) & $m=1$ & $m=2$ & $m=3$   \\
\hline
1 & $-2.0\times 10^{-6}$ & $48.9$ ($50.3$)  & $-140$ ($-148$) & $-1.5$ ($-1.8$) &  $0.7$ & $0.5$ & $0.5$   \\
2 & $-5.0\times 10^{-6}$ & $74.2$ ($75.2$)  & $-348$ ($-377$) & $-3.0$  ($-3.4$) &  $1.1$ & $0.7$ & $0.6$ \\
3 & $-1.0\times 10^{-5}$ & $98.2$ ($105.3$)  & $-599$ ($-658$) & $-4.5$ ($-5.3$) &  $5.3$ & $0.6$ & $0.5$  \\
4 & $-2.0\times 10^{-5}$ & $127.3$ ($137.3$) & $-937$ ($-1025$) & $-6.6$ ($-7.7$) &  $9.5$ & $1.0$ & $0.8$  \\
5 & $-2.2\times 10^{-5}$ & $131.3$ ($141.8$) & $-981$ ($-1080$) & $-6.8$ ($-8.1$) &  $13.1$ & $1.0$ & $0.8$  \\
6 & $-2.4\times 10^{-5}$ & $134.7$ ($145.9$) & $-1022$ ($-1131$) & $-7.0$ ($-8.4$) &  $14.5$ & $1.3$ & $0.8$  \\
7 & $-2.6\times 10^{-5}$ & $109.5$ ($149.6$) & $-725$ ($-1178$) & $-5.1$ ($-8.7$) &  $33.1$ & $8.3$ & $2.2$  \\
8 & $-3.0\times 10^{-5}$ & $100.2$ ($156.3$) & $-706$ ($-1264$) & $-4.9$ ($-9.2$) &  $35.5$ & $10.0$ & $2.9$ \\
9 & $-4.0\times 10^{-5}$ & $91.2$ ($169.2$)  & $-698$ ($-1435$) & $-4.6$ ($-10.3$) &  $39.3$ & $12.7$ & $4.5$ \\
10& $-6.0\times 10^{-5}$ & $85.4$ ($185.1$)  & $-708$ ($-1662$) & $-4.2$ ($-11.6$) &  $44.6$ & $15.6$ & $6.3$ \\
\hdashline[1pt/1pt]
Observed &  $-$ & $130.8 \pm 4.5$ & $-$ & $-$ & $11.8\pm 2.4$ & $3.8\pm 0.6$ & $1.9\pm 0.7$ \\
\hline
\end{tabular}
\label{table:1}
\end{center}
\end{table}
\renewcommand{\arraystretch}{1.0}

\newpage
\clearpage


\setcounter{figure}{0}
\setcounter{table}{0}
\setcounter{equation}{0}
\renewcommand{\thefigure}{S\arabic{figure}}
\renewcommand{\theequation}{S\arabic{equation}}
\renewcommand{\thetable}{S\arabic{table}}


\newpage
\clearpage

\begin{center}
{\LARGE
Supplementary Materials:\\ The Sun's differential rotation is controlled by high-latitude baroclinically unstable inertial modes
}
\\
\vspace{0.2cm}
{\large Yuto~Bekki,$^{1}$ Robert~H.~Cameron,$^{1}$ Laurent~Gizon$^{1,2,3,*}$ }
\\
\vspace{0.4cm}
{$^{1}$Max-Planck-Institut f\"ur Sonnensystemforschung,  37077 G\"ottingen, Germany \\
$^{2}$Institut f\"ur Astrophysik, Georg-August-Universit\"at G\"ottingen, 37077 G\"ottingen, Germany \\$^{3}$Center for Space Science,  New York University Abu Dhabi, Abu Dhabi, UAE
}
\\
{$^{\ast}$Corresponding author. Email:  gizon@mps.mpg.de}
\\
\vspace{0.6cm}
\end{center}

\begin{figure}[]
\begin{center}
\includegraphics[scale=0.65]{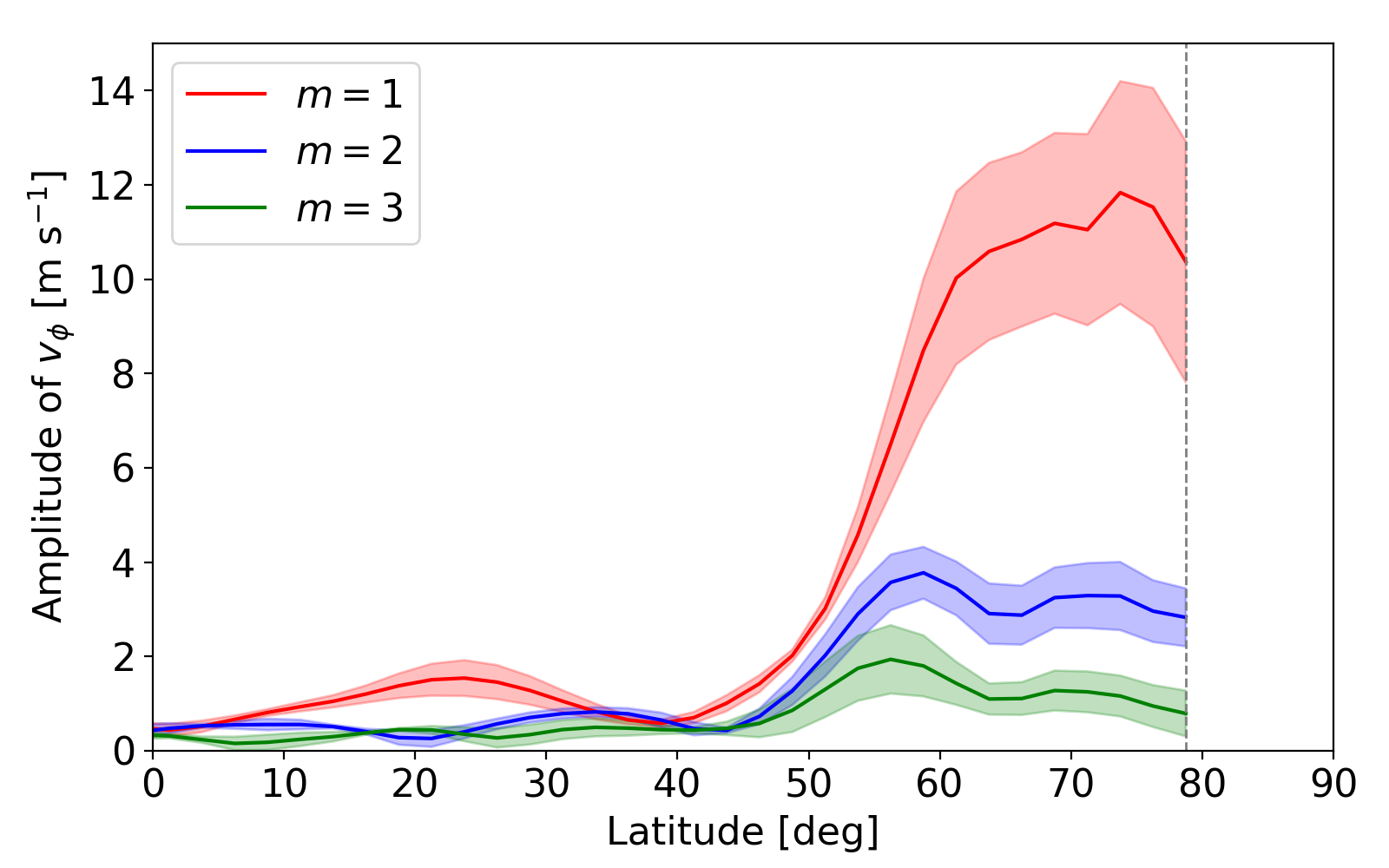}
\caption{
{{\bf Maximum amplitude of the longitudinal velocity of the observed high-latitude inertial modes near the solar surface for $m=1$, $2$, and $3$.}
The velocities are derived from ring-diagram helioseismology using $5^{\circ}$ tiles \cite{gizon2021} from  SDO/HMI data from 2017-2021.
The shaded areas indicate the $1\sigma$ error estimates.}
\label{fig_s:vobs_m123}}
\end{center}
\end{figure}

\begin{itemize}
\item[-] Supplementary Text
\item[-] Figures~\ref{fig_s:vobs_m123}--\ref{fig_s:fig4_lambda}
\item[-] Table~\ref{table_s:lambda}
\end{itemize}

\clearpage
\newpage
\section*{Thermal wind balance approximation} 

Let us consider the longitudinally-averaged equation of motion in the Sun's convection zone in a frame rotating at an angular velocity $\bm{\Omega}_{0}=\Omega_{0}\bm{e}_{z}$.
We assume that the large-scale mean flows (differential rotation and meridional circulation) are in a statistically stationary state and the dominant force balance is achieved by the pressure gradient force, the buoyancy force, and the Coriolis force, i.e., the Reynolds stress, Maxwell stress (Lorentz force), and viscous diffusion are all assumed to be small compared to the Coriolis force.
Under these assumptions, the equation of motion can be reduced to 
\begin{eqnarray}
&& 0\approx \frac{\nabla \overline{p}_{1}}{\rho_{0}} +\frac{\overline{\rho}_{1}}{\rho_{0}} g \bm{e}_{r}+2\bm{\Omega}_{0}\times \overline{\bm{v}}. \label{eq_s:twb}
\end{eqnarray}
where the overbar represents the longitudinal averages, $p_{0}(r)$ and $\rho_{0}(r)$ denote the radial functions of background pressure and density in a hydrostatic equilibrium with the gravitational acceleration $g(r)>0$, whereas $p_{1}(r,\theta,\phi)$ and $\rho_{1}(r,\theta,\phi)$ are the perturbations of pressure and density with respect to the background stratification.
The mean velocity $\overline{\bm{v}}$ is the sum of the the meridional circulation $\overline{\bm{v}}_{m}=\overline{v}_{r}\bm{e}_{r}+\overline{v}_{\theta}\bm{e}_{\theta}$ and the differential rotation $\overline{v}_{\phi}\bm{e}_{\phi}=r\sin{\theta}{\Omega}_{1}\bm{e}_{\phi}$.
In the following discussion, we use the differential rotation profile ${\Omega}_{1}(r,\theta)$ determined by global helioseismology \cite{howe2005}.

The $\theta$-component of eq.~(\ref{eq_s:twb}) can be written as
\begin{eqnarray}
&& \frac{\partial\overline{p}_{1}}{\partial \theta} \approx 2r^{2}\rho_{0}\Omega_{0}{\Omega}_{1} \sin{\theta}\cos{\theta}. \label{eq_s:dpdq}
\end{eqnarray}
This balance between the pressure gradient force and the Coriolis force is called geostrophic balance.
The mean pressure perturbation $\overline{p}_{1}$ can be estimated by integrating eq.~(\ref{eq_s:dpdq}) 
\begin{eqnarray}
\overline{p}_{1}(r,\theta) \approx \int_{0}^{\theta} 2r^{2}\rho_{0}\Omega_{0}{\Omega}_{1}(r,\theta^{\prime}) \sin{\theta^{\prime}}\cos{\theta^{\prime}} d\theta^{\prime} + \mathcal{C}_{p}(r), \label{eq_s:p1_sun}
\end{eqnarray}
where the radial function $\mathcal{C}_{p}(r)$ can be set by $\int_{0}^{\pi}\overline{p}_{1} \sin{\theta}d\theta =0$ at each height.

By taking a curl of the eq.~(\ref{eq_s:twb}), we obtain the equation of \textit{thermal wind balance}
\begin{eqnarray}
&& \frac{\partial\overline{s}_{1}}{\partial \theta} \approx \frac{2c_{\mathrm{p}}}{g} r^{2}\sin{\theta}\Omega_{0}\frac{\partial {\Omega}_{1}}{\partial z}, \label{eq_s:dsdq}
\end{eqnarray}
where $c_{\mathrm{p}}=4.17\times 10^{8}$ erg g$^{-1}$ K$^{-1}$ is the specific heat at constant pressure and $\partial/\partial z =\cos{\theta} \partial/\partial r- r^{-1}\sin{\theta} \partial /\partial\theta$.
Equation~(\ref{eq_s:dsdq}) can be used to estimate the entropy perturbation in the Sun's convection zone as
\begin{eqnarray}
&& \overline{s}_{1}(r,\theta) \approx \int_{0}^{\theta} \frac{2c_{\mathrm{p}}}{g} r^{2}\sin{\theta^{\prime}}\Omega_{0}\frac{\partial {\Omega}_{1}(r,\theta^{\prime})}{\partial z} d\theta^{\prime} + \mathcal{C}_{s}(r). \label{eq_s:s1_sun}
\end{eqnarray}
Here, the integral constant $\mathcal{C}_{s}(r)$ is determined by $\int_{0}^{\pi}\overline{s}_{1} \sin{\theta}d\theta =0$ at each height.
From the linearized equation of state, the temperature perturbation in the Sun can be estimated as follows
\begin{eqnarray}
&& \overline{T}_{1} = T_{0} \left[ \frac{\overline{s}_{1}}{c_{\mathrm{p}}} -\frac{\gamma-1}{\gamma} \frac{\overline{p}_{1}}{p_{0}} \right], \label{eq_s:T1_sun}
\end{eqnarray}
where $\gamma=c_{\mathrm{p}}/c_{\mathrm{v}}=5/3$ denotes the specific heat ratio.
The profiles of the solar differential rotation and the estimated $\overline{s}_{1}$ and $\overline{T}_{1}$ are shown in Fig.~\ref{fig:1}B and C.

\section*{Linear stability analysis}  \label{sec:linana}

We study the linear stability of the high-latitude inertial modes by solving a two-dimensional (2D) eigenvalue problem of the rotating fluid with a realistic solar convection zone model \cite{bekki2022a}.
We numerically solve the linearized equation of continuity, the equation of motion, the equation of entropy, and the equation of state in a spherical coordinate ($r,\theta,\phi$).
The model includes the solar background stratification (adiabatic), spatially-uniform turbulent viscous and thermal diffusivities of $10^{12}$ cm$^{2}$ s$^{-1}$, and the solar differential rotation from global helioseismology. 
For simplicity, we assume that the stratification is adiabatic in the radial direction.
We include a background latitudinal entropy variation throughout the convection zone of the following simple form,
\begin{eqnarray}
&& \frac{\partial s} {\partial\theta}=\Delta_{\theta}\overline{s} \sin{(2\theta)},\label{eq_s:twb_sin}
\end{eqnarray}
where $\Delta_{\theta} \overline{s} = \overline{s}_{\mathrm{eq}} - \overline{s}_{\mathrm{pole}}$ denotes the (negative) entropy difference between the cooler equator and the hotter poles and is assumed to be radially-uniform.
We vary $\Delta_{\theta} \overline{s}$ from $0$ to $-2000$ erg g$^{-1}$ K$^{-1}$ as a free parameter.
See Section 6.2 in Ref.~\cite{bekki2022a} for more details.

The numerical domain extends from the base of the convection zone at $r=0.71R_{\odot}$ to the top boundary at $r=0.985R_{\odot}$.
We solve the linearized equations assuming that the perturbations are proportional to $\propto \exp(\ii m\phi-\ii \omega t)$, where $m$ is the azimuthal order and $\omega$ is the angular frequency.
The spatial derivatives are evaluated using the second-order finite-difference method.
An impenetrable and stress-free boundary condition is used at the top and bottom boundaries.
On the polar axis, we assume that all the variables ($\rho_{1}$, $v_{r}$, $v_{\theta}$, $v_{\phi}$, and $s_{1}$) vanish for $m>1$. 
For $m=1$, we use the special boundary condition $\partial v_{\theta}/\partial\theta =0$ to allow a flow through the polar axis.
At each $m$, we select the fastest-growing (or least-damped) high-latitude mode with north-south antisymmetric longitudinal velocity $v_{\phi}$.

\begin{figure}
\begin{center}
\vspace{-1.3cm}
\includegraphics[scale=0.57]{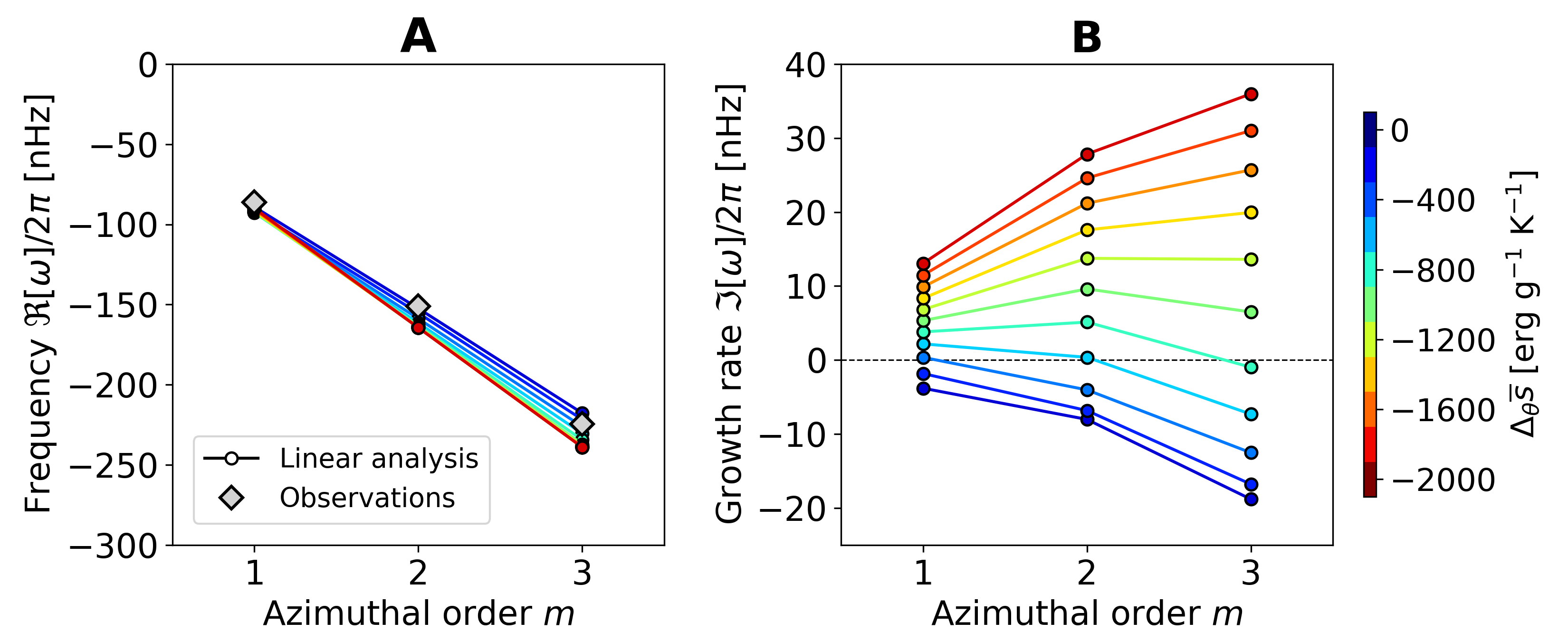}
\caption{ 
{\bf Eigenfrequencies of the high-latitude inertial modes with north-south antisymmetric $v_{\phi}$ for $1 \leq m \leq 3$.}
(A) Real part of the eigenfrequencies, $\Re[\omega]/2\pi$, obtained from the linear analysis in the Carrington reference frame.
Positive (negative) $\Re[\omega]$ indicates prograde (retrograde) propagation.
The different colors represent the results for different values of the imposed latitudinal entropy variation $|\Delta_{\theta} \overline{s}|$.
The gray diamonds show the observed angular frequencies of the high-latitude inertial modes for $1 \leq m \leq 3$ \cite{gizon2021}.
(B) The growth rates of the high-latitude modes.
Positive (negative) $\Im[\omega]$ indicates that the mode is growing (decaying).
}
\label{fig_s:linana_dispersion}
\end{center}
\end{figure}

The frequencies $\Re[\omega]$ and the growth rates $\Im[\omega]$ of the high-latitude inertial modes are shown in fig.~\ref{fig_s:linana_dispersion} for azimuthal orders $1 \leq m \leq 3$. 
The observed frequencies of the low-$m$ high-latitude modes can be nicely reproduced by the linear dispersion relation which is almost independent of $\Delta_{\theta}  \overline{s} $ (fig.~\ref{fig_s:linana_dispersion}A).
On the other hand, fig.~\ref{fig_s:linana_dispersion}B shows that their growth rates increase as $\Delta_{\theta}  \overline{s} $ decreases, i.e., the modes become more and more linearly unstable as the convection zone becomes more and more baroclinic.
Since the background stratification is set to be convectively neutral, we refer to these modes as baroclinically-unstable modes (or baroclinic modes).
We note that  the $m=1$ mode is the first mode that becomes unstable as $|\Delta_{\theta}  \overline{s} |$ increases (when $|\Delta_{\theta} \overline{s}| \approx 500$ erg g$^{-1}$ K$^{-1}$).
For sufficiently large baroclinicity ($|\Delta_{\theta} \overline{s}| \gtrsim 1200$ erg g$^{-1}$ K$^{-1}$), the modes with $m=1,2,3$ become all unstable.

\begin{figure}
\begin{center}
\includegraphics[scale=0.23]{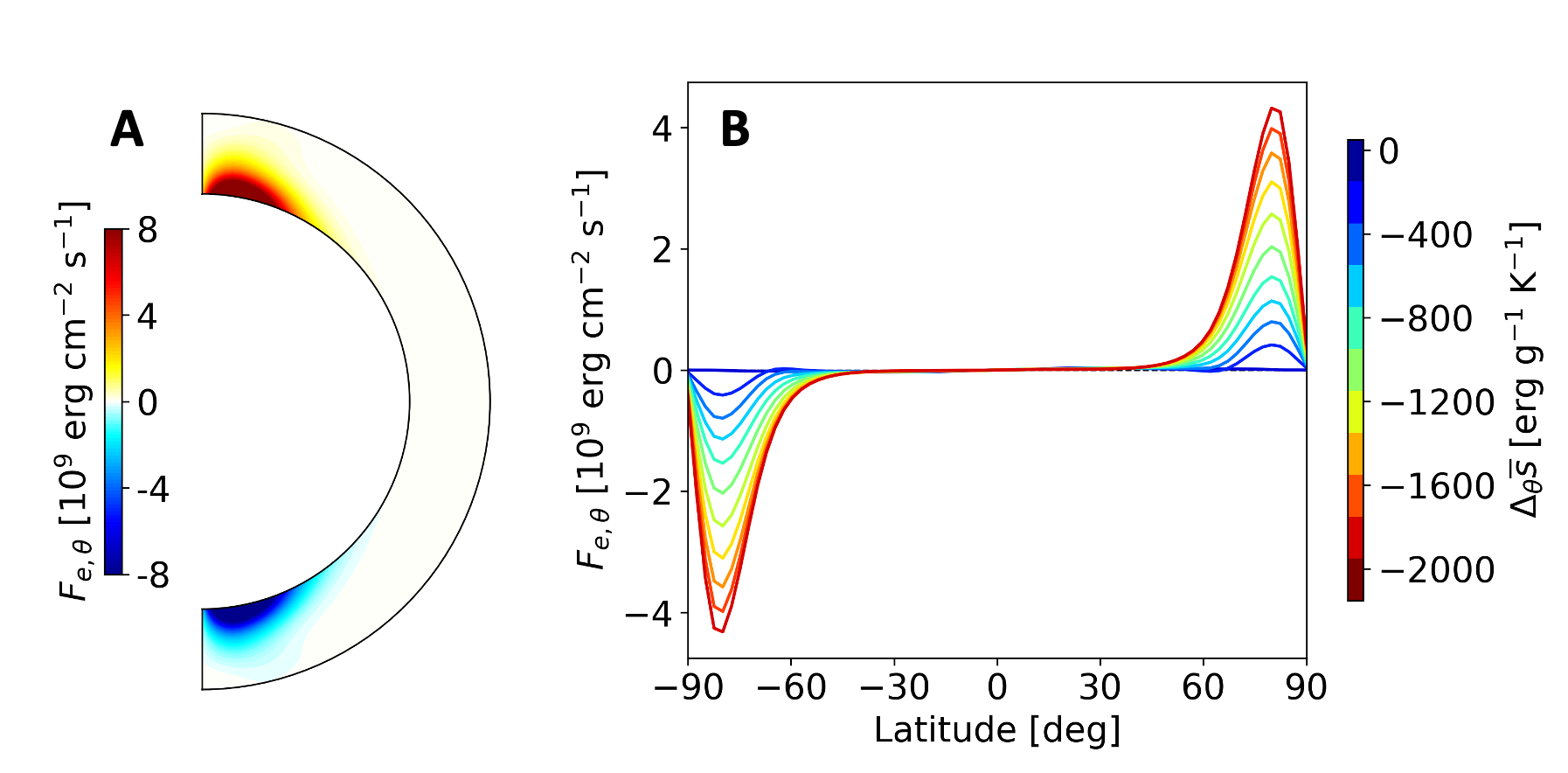}
\caption{
{\bf Latitudinal component of the thermal energy flux $F_{e,\theta}=\rho_{0} c_{\mathrm{p}}\overline{ v_{\theta} T_{1} }$ associated with the $m=1$ high-latitude mode obtained from the linear analysis.}
(A) The meridional plot of $F_{e,\theta}$ for the case with $\Delta_{\theta}\overline{s}=-2000$ erg g$^{-1}$ K$^{-1}$.
(B) Radially-averaged profiles of $F_{e,\theta}$ with varying $\Delta_{\theta} \overline{s}$.
In all cases, the eigenfunctions are normalized such that the maximum velocity amplitude of $v_{\phi}$ is $11.8$ m~s$^{-1}$ at the surface, as suggested by the observation.
}
\label{fig_s:linana_Fe}
\end{center}
\end{figure}

We find that these linear baroclinic modes have a general tendency to transport  heat equatorward at high latitudes.
This is seen in fig.~\ref{fig_s:linana_Fe}A which shows the latitudinal enthalpy flux $F_{e,\theta}=\rho_{0} c_{\mathrm{p}}\overline{ v_{\theta} T_{1} }$ of the $m=1$ mode.
It is also shown that the amount of equatorward heat transport is enhanced as $|\Delta_{\theta} \overline{s}|$ increases and the convection zone becomes more baroclinic (fig.~\ref{fig_s:linana_Fe}B).
Therefore, the baroclinic modes are expected to have a substantial impact on the background baroclinicity in the nonlinear regime.

\section*{Numerical model: 3D nonlinear simulations}

In order to study the baroclinic excitation and the nonlinear saturation of the high-latitude modes, we carry out a set of three-dimensional (3D) hydrodynamic simulations of the large-scale flows in the Sun.

In rotating convection simulations \cite{bekki2022b}, the observed properties of the high-latitude inertial modes are not properly reproduced because the simulations do not have a differential rotation profile (or the associated latitudinal entropy variation) which is close to that of the Sun.
This is related to the convective conundrum \cite{hotta2022,hotta2023}.
In this study, we therefore use a mean-field approach where the small-scale turbulent convection is not solved but instead the temporally-averaged convective angular momentum transport ($\Lambda$ effect) is parameterized in the model \cite{rudiger1989}.
This enables us to study the high-latitude modes in the context of a solar-like differential rotation profile.

Our equations are nonlinear, so as the baroclinically-unstable modes grow in time, they extract the available potential energy from the background latitudinal entropy difference.
In the Sun, this potential energy is possibly replenished either by the latitudinal convective heat transport \cite{hotta2018,kitchatinov1995,kuker2001} or by radial penetration of meridional circulation into the weakly-subadiabatic overshooting layer \cite{rempel2005,brun2011}.
In this study, we adopt the latter mechanism and include the weakly-subadiabatic layer near the base of the convection zone to sustain the latitudinal entropy gradient.
In order to vary the baroclinicity, we change the subadiabaticity in the overshooting layer which is a free parameter of the model.

We numerically solve the 3D mean-field hydrodynamic equations in spherical geometry $(r,\theta,\phi)$.
The equations are 
\begin{eqnarray}
&& \frac{\partial \rho_{1}}{\partial t} = -\frac{1}{\xi^{2}}\nabla\cdot (\rho_{0} \bm{v}), \label{eq_s:mass2} \\
&& \frac{\partial \bm{v}}{\partial t} = -\bm{v}\cdot\nabla\bm{v}-\frac{\nabla p_{1}}{\rho_{0}}-\frac{\rho_{1}}{\rho_{0}}g\bm{e}_{r}+2\bm{v}\times\Omega_{0}\bm{e}_{z}+\frac{1}{\rho_{0}}\nabla\cdot \bm{\mathcal{R}}, \label{eq_s:motion2} \\
&& \frac{\partial s_{1}}{\partial t} = -\bm{v}\cdot\nabla s_{1}+c_{\mathrm{p}}\delta\frac{v_{r}}{H_{p}}
    +\frac{1}{\rho_{0}T_{0}}\nabla\cdot (\rho_{0}T_{0}\kappa\nabla s_{1}) +\frac{1}{\rho_{0}T_{0}}(\bm{\mathcal{R}}\cdot\nabla)\cdot\bm{v}, \label{eq_s:entropy2} 
\end{eqnarray}
where $\bm{v}$, $p_{1}$, $\rho_{1}$, and $s_{1}$ are the velocity, pressure perturbation, density perturbation, and entropy perturbation from the hydrostatic equilibrium background state (quantities with subscript $0$).
We use the same background stratification as in Ref.~\cite{rempel2005}:
\begin{eqnarray}
&& g(r)=g_{\mathrm{bc}} \left( \frac{r}{r_{\mathrm{bc}}}\right)^{-2}, \\
&& T_{0}(r)=T_{\mathrm{bc}} \left[1+\frac{\gamma-1}{\gamma} \frac{r_{\mathrm{bc}}}{H_{\mathrm{bc}}} \left( \frac{r_{\mathrm{bc}}}{r}-1\right) \right], \\
&& \rho_{0}(r)=\rho_{\mathrm{bc}} \left[1+\frac{\gamma-1}{\gamma} \frac{r_{\mathrm{bc}}}{H_{\mathrm{bc}}} \left( \frac{r_{\mathrm{bc}}}{r}-1\right) \right]^{1/(\gamma-1)}, \\
&& p_{0}(r)=p_{\mathrm{bc}} \left[1+\frac{\gamma-1}{\gamma} \frac{r_{\mathrm{bc}}}{H_{\mathrm{bc}}} \left( \frac{r_{\mathrm{bc}}}{r}-1\right) \right]^{\gamma/(\gamma-1)}.
\end{eqnarray}
Here, $g_{\mathrm{bc}}=5.3\times 10^{4}$~cm~s$^{-2}$, $T_{\mathrm{bc}}=2.2\times 10^{6}$~K, $\rho_{\mathrm{bc}}=0.202$~g~cm$^{-3}$, $p_{\mathrm{bc}}=6.1 \times 10^{13}$~dyn~cm$^{-2}$, and $H_{\mathrm{bc}}=p_{\mathrm{bc}}/(\rho_{\mathrm{bc}}g_{\mathrm{bc}})=0.0827R_{\odot}$ are the values of gravitational acceleration, temperature, density, pressure, and pressure scale height at the base of the convection zone $r_{\mathrm{bc}}=0.71R_{\odot}$.
We use the linearized equation of state assuming a perfect gas consisting of fully-ionized hydrogen,
\begin{eqnarray}
&& p_{1}=p_{0}\left( \gamma\frac{\rho_{1}}{\rho_{0}} +\frac{s_{1}}{c_{\mathrm{v}}}\right), \label{eq_s:state2}
\end{eqnarray}
where $\gamma=c_{\mathrm{p}}/c_{\mathrm{v}}=5/3$ is the ratio of specific heats.

\begin{figure}
\begin{center}
\vspace{-1.5cm}
\includegraphics[scale=0.625]{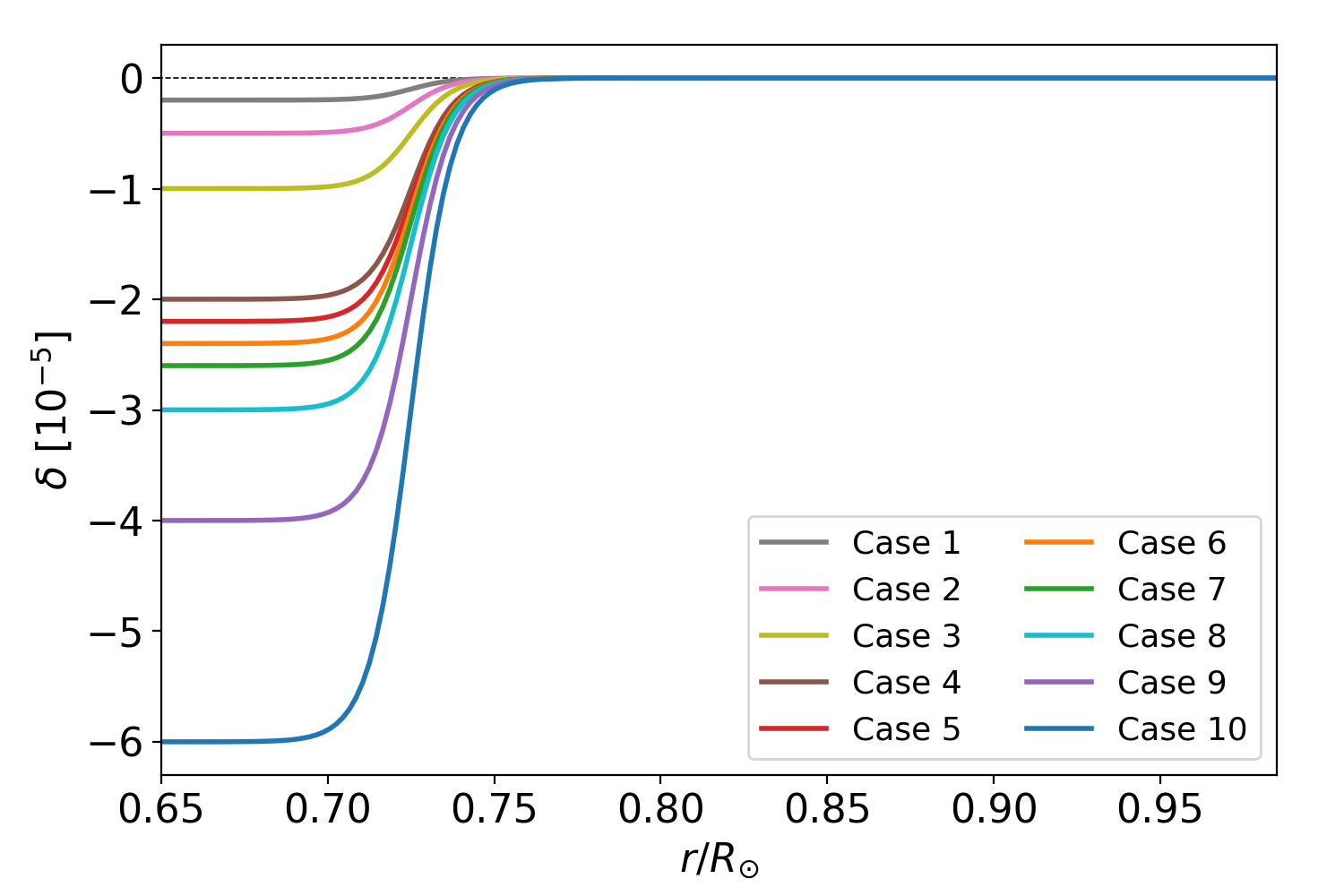}
\caption{ 
{\bf Radial profiles of the superadiabaticity $\delta(r)$ used in our nonlinear simulations.}
\label{fig_s:delta_nonlin}
  }
\end{center}
\end{figure}

In accordance with previous works \cite{rempel2005},
the superadiabaticity $\delta=\nabla-\nabla_{\mathrm{ad}}$ with $\nabla=d\ln{T}/d\ln{p}$ is chosen  to be 
\begin{eqnarray}
&& \delta(r)=\frac{\delta_{0}}{2}\left[1-\tanh{\left(\frac{r-r_{\mathrm{sub}}}{d_{\mathrm{sub}}} \right)}\right],
\end{eqnarray}
where $r_{\mathrm{sub}}=0.725R_{\odot}$ and $d_{\mathrm{sub}}=0.0125R_{\odot}$.
With this formulation, $\delta_{0}$ is zero throughout most of the convection zone (because the effects of convection are included in a parameterized form using the mean-field framework) and $\delta_{0}<0$ only below the base of the convection zone.
The parameter $\delta_{0}$ then can be used to control the baroclinicity via 
eq.~(\ref{eq_s:entropy2}).
A non-zero $\delta_0$ introduces entropy fluctuations at the base of the convection zone owing to the meridional circulation that extends down to the overshooting layer.
The mean baroclinicity is established in the convection zone as the entropy perturbations spread into the convection zone above by turbulent diffusion, see Ref.~\cite{rempel2005}. 
In this study, we vary $\delta_0$ from $-2\times 10^{-6}$ to $-6\times 10^{-5}$ {as shown in fig.~\ref{fig_s:delta_nonlin}}.

The Reynolds stress tensor $\bm{\mathcal{R}}$ is decomposed into the turbulent diffusive part and the non-diffusive part ($\Lambda$ effect) as,
\begin{eqnarray}
&& \mathcal{R}_{ik}=\rho_{0}\nu \left[\left(\mathcal{S}_{ik}
    -\frac{2}{3}\delta_{ik} \nabla\cdot\bm{v} \right) 
    +\Lambda_{ik}\Omega_{0} \right], \label{eq_s:reynolds}
\end{eqnarray}
where $S_{ik}$ and $\delta_{ik}$ denote the velocity deformation tensor and Kronecker-delta unit tensor.
The dimensionless tensor $\Lambda_{ik}$ specifies the $\Lambda$ effect.
We use a functional form of the $\Lambda$ effect similar to that of a previous study \cite{rempel2005}:
\begin{eqnarray}
\Lambda_{r\phi}&=&\Lambda_{0} \tilde{f}_{\Lambda}(r,\theta) \cos{(\theta+\lambda)} \left[ 1+\sigma_{r}(r,\theta,\phi)\right], 
 \label{eq.lambda-rp}
 \\
 \Lambda_{\theta \phi}&=&-\Lambda_{0} \tilde{f}_{\Lambda}(r,\theta) \sin{(\theta+\lambda)} \left[ 1+\sigma_{\theta}(r,\theta,\phi)\right],
 \label{eq.lambda-tp}
\end{eqnarray}
where $\Lambda_{0}$ is a dimensionless parameter which determines the overall amplitude of the $\Lambda$ effect.
In this study, we use a fixed value $\Lambda_{0}=0.85$ (except for additional simulations reported in the supplementary section "Dependence on $\Lambda$ effect").
The spatial distribution of the $\Lambda$ effect in the above equations is specified by
\begin{eqnarray}
&& \tilde{f}_{\Lambda}(r,\theta)=\frac{f_{\Lambda}(r,\theta)}{\mathrm{max}|f_{\Lambda}(r,\theta)|}, \\
&& f_{\Lambda}(r,\theta)=\sin^{2}{\theta}\cos{\theta} \tanh{\left(\frac{r_{\mathrm{max}}-r}{d_{\mathrm{sf}}} \right)},
\end{eqnarray}
where $d_{\mathrm{sf}}=0.025R_{\odot}$.
The inclination angle of the $\Lambda$ effect in eqs.~(\ref{eq.lambda-rp}) and~(\ref{eq.lambda-tp}) is fixed to be $\lambda=15^{\circ}$ in the northern hemisphere and $-15^{\circ}$ in the southern hemisphere.
Therefore, the angular momentum is dominantly transported equatorward and weakly transported cylindrically outward by the $\Lambda$~effect.
This parameterization ensures that the resulting differential rotation is solar-like (with faster equator and slower poles) and the resulting meridional circulation is largely single-cell in each hemisphere, as suggested by recent helioseismic observations \cite{gizon2020s}.
To excite the non-axisymmetric modes, we follow the method described in Ref.~\cite{bekki2023} and add random fluctuations ($\sigma_{r}$ and $\sigma_{\theta}$) in the $\Lambda$ effect which mimic the stochastic convective motions.
As for the turbulent viscosity $\nu$ and the thermal diffusivity $\kappa$, we use the same functional forms as in Ref.~\cite{rempel2005}.

\section*{Non-axisymmetric modes}

\begin{figure}
\begin{center}
\includegraphics[scale=0.65]{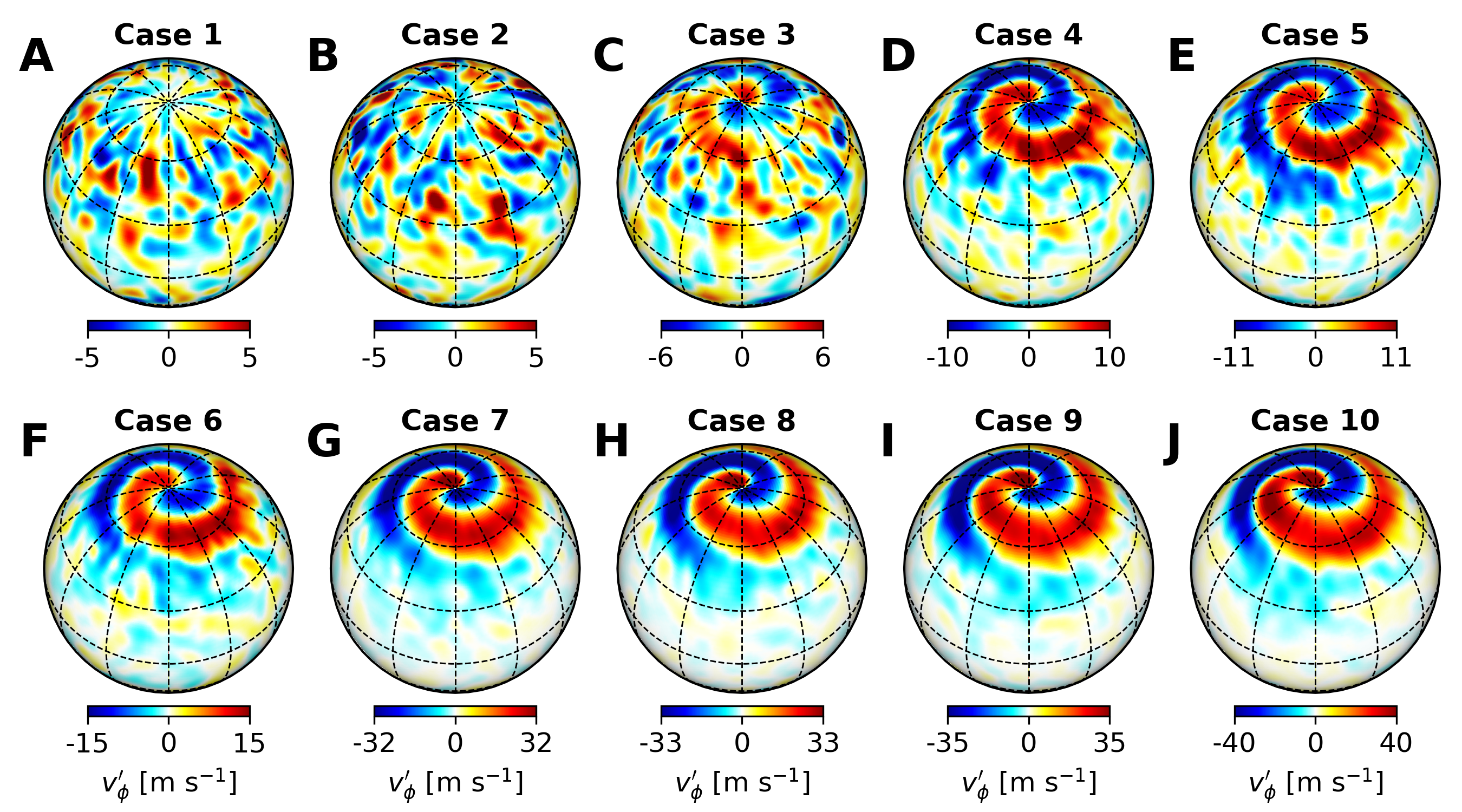}
\caption{ 
{\bf Snapshots of the non-axisymmetric component of the longitudinal velocity $v_{\phi}^{\prime}$ at the top of the simulations ($r=0.985R_{\odot}$)}.
Panels A--J show the results from simulation cases 1--10 of Table~\ref{table:1}.
The results are shown at $t=40$~yr when the statistically stationary solutions are reached.
\label{fig_s:final_vphi}
  }
\end{center}
\end{figure}

\begin{figure}[t!]
\begin{center}
\vspace{-1.5cm}
\includegraphics[scale=0.65]{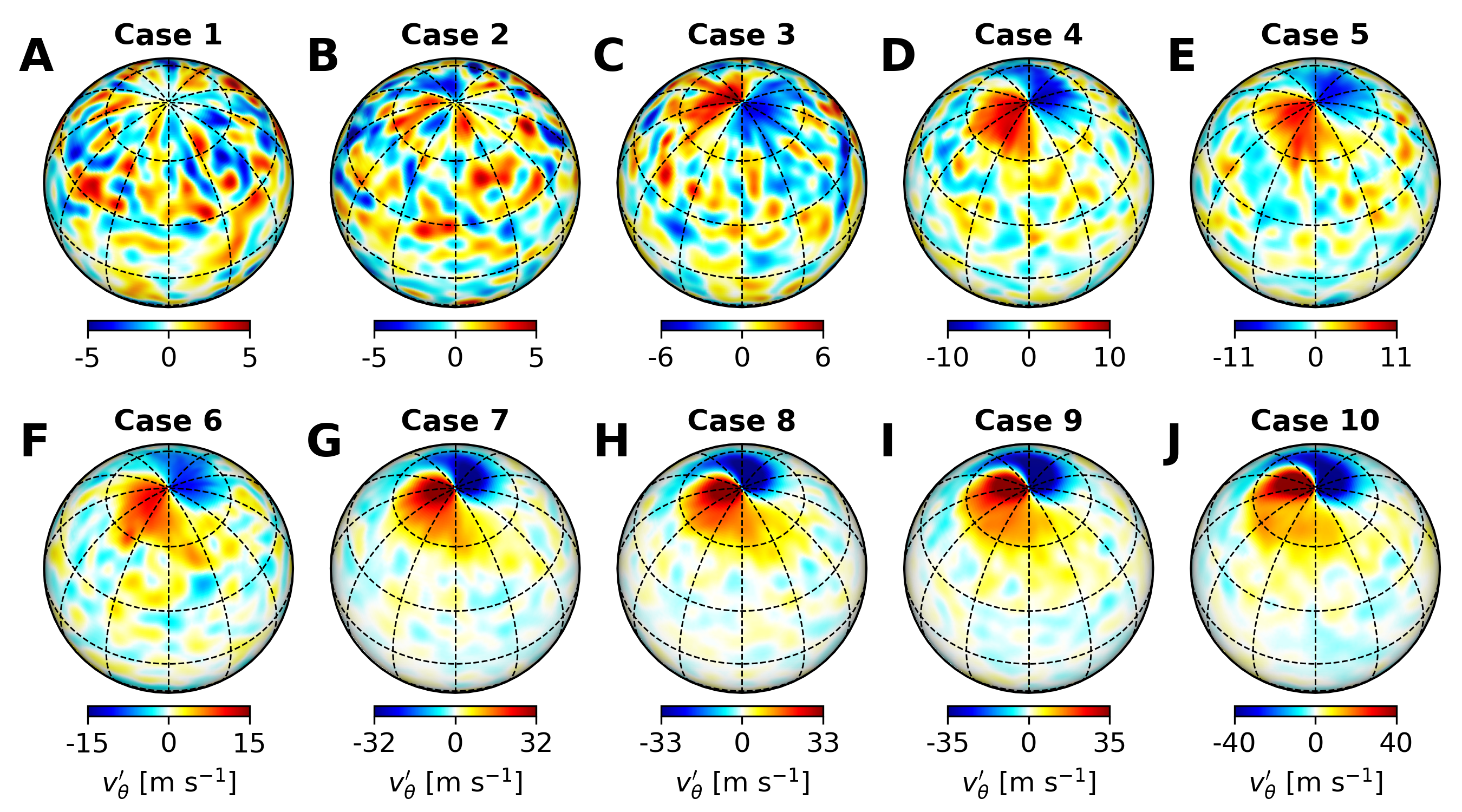}
\caption{ 
{\bf Snapshots of the non-axisymmetric component of the latitudinal velocity $v_{\theta}^{\prime}$ at the top of the simulations}.
Panels A--J show the results from simulation cases 1--10 of Table~\ref{table:1}.
The results are shown at $t=40$~yr when the statistically stationary solutions are reached.
\label{fig_s:final_vtheta}
  }
\end{center}
\end{figure}

\begin{figure}[b!]
\begin{center}
\vspace{-1.5cm}
\includegraphics[scale=0.65]{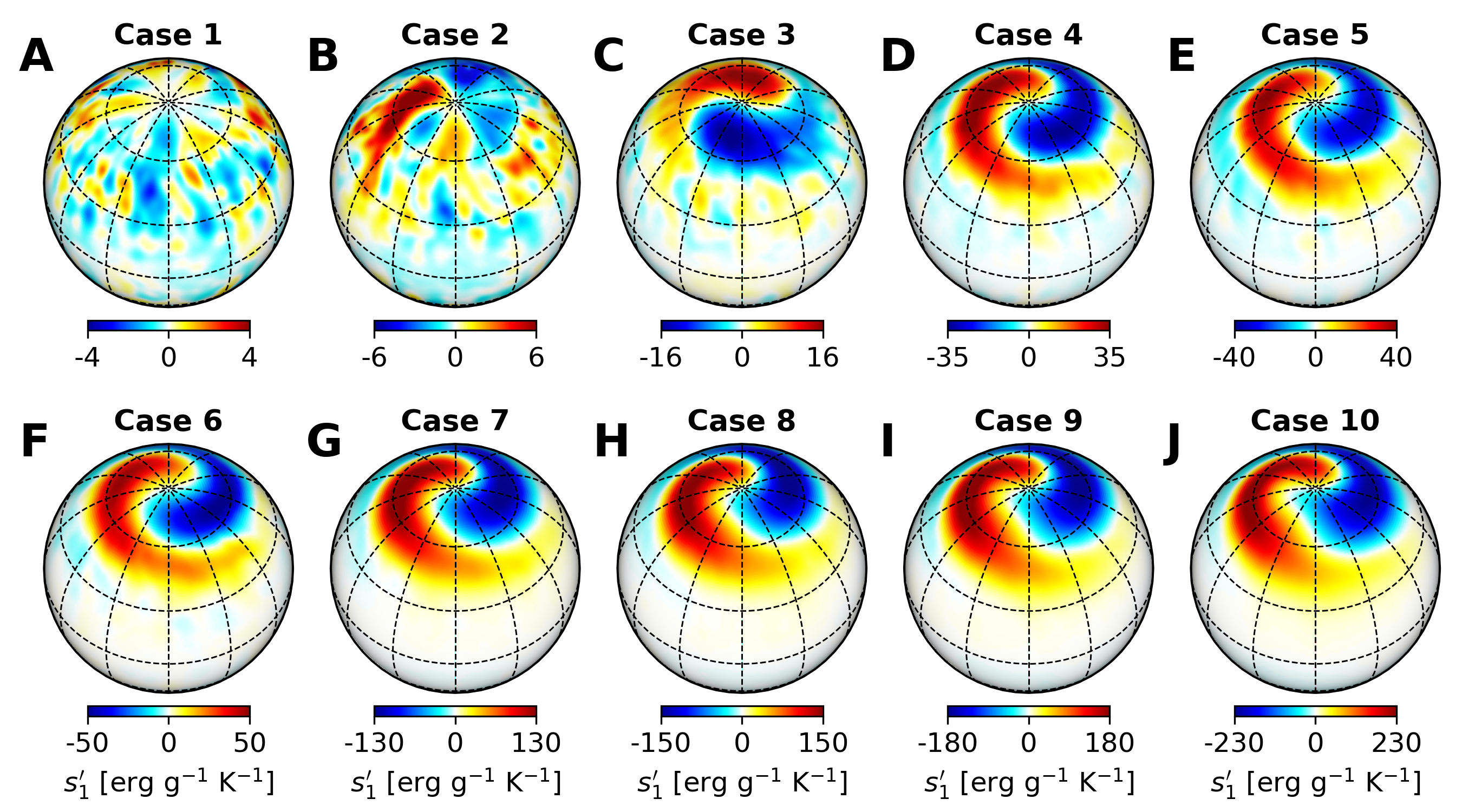}
\caption{ 
{\bf Snapshots of the non-axisymmetric component of the entropy perturbation $s_{1}^{\prime}$ at the top of the simulations}.
Panels A--J show the results from simulation cases 1--10 of Table~\ref{table:1}.
The results are shown at $t=40$~yr when the statistically stationary solutions are reached.
\label{fig_s:final_s1}
  }
\end{center}
\end{figure}

\begin{figure}
\begin{center}
\includegraphics[scale=0.72]{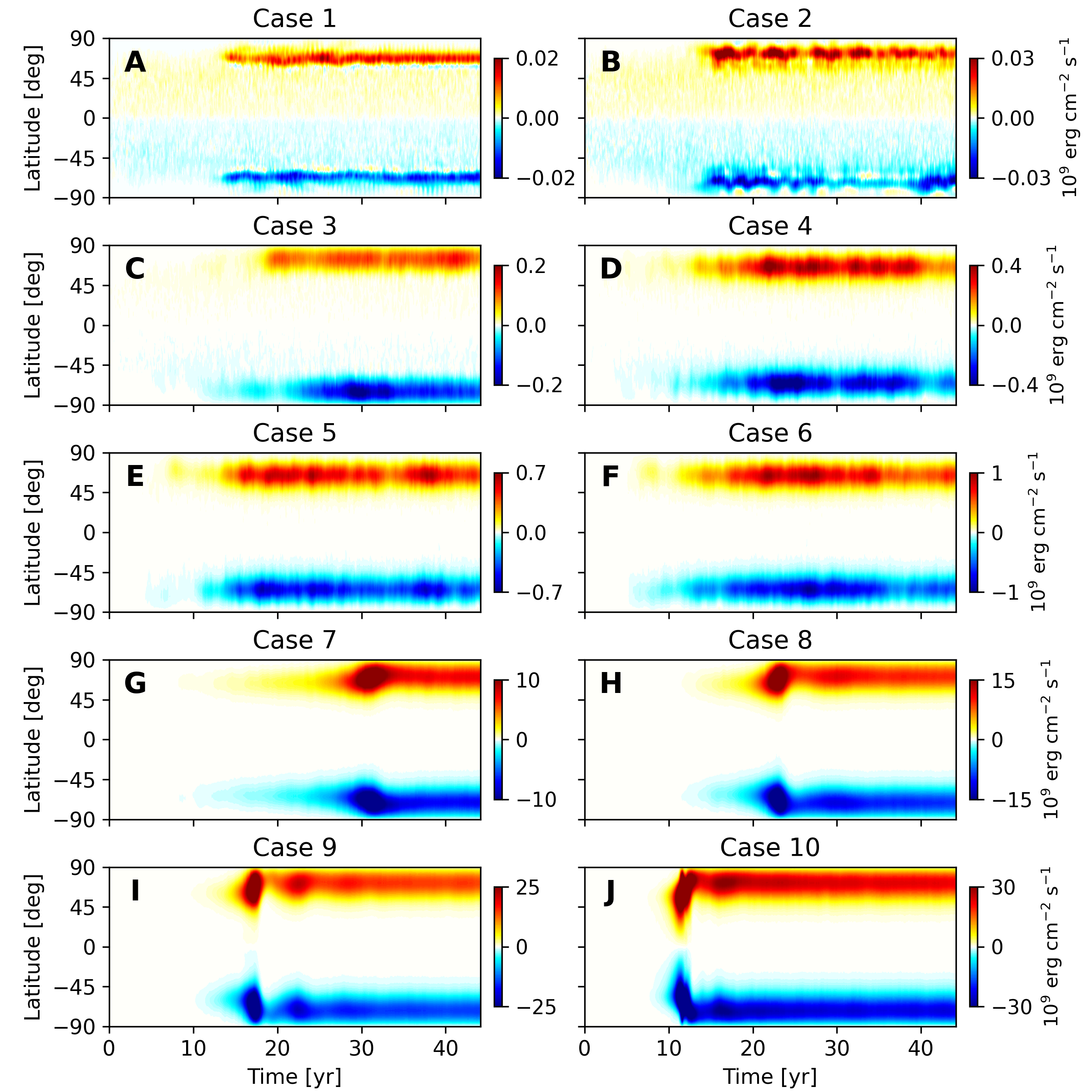}
\caption{ 
{\bf Latitudinal heat transport by non-axisymmetric modes.}
Time-latitude plots of the latitudinal component of heat flux $\rho_{0}T_{0} \overline{v_{\theta}^{\prime} s_{1}^{\prime}}$ averaged in radius over the convection zone ($0.71R_{\odot}-0.985R_{\odot}$).
Panels A--J show the results from cases 1--10.
\label{fig_s:Feq}}
\end{center}
\end{figure}

The simulations are run for about $45$ years (simulated time).
As shown in Fig.~\ref{fig:2}, a statistically-stationary state is reached by about $30$ years.
Figure~\ref{fig_s:final_vphi} shows temporal snapshots of the non-axisymmetric component of longitudinal velocity $v_{\phi}^{\prime}$ at the surface $r=0.985R_{\odot}$ in the statistically-stationary state for the different cases.
Here, the superscript $^{\prime}$ denoting the perturbation with respect to its longitudinal mean.
From case 1 to case 10, the velocity amplitudes of the high-latitude modes increase.
The $m=1$ mode is predominant in cases 3--10.
Figures~\ref{fig_s:final_vtheta} and \ref{fig_s:final_s1} show the same snapshots as fig.~\ref{fig_s:final_vphi} but for the non-axisymmetric components of the latitudinal velocity $v_{\theta}^{\prime}$ and the entropy perturbation $s_{1}^{\prime}$ at the surface.

Figure~\ref{fig_s:Feq} shows the time-latitude plots of the radially-averaged latitudinal heat flux $\rho_{0}T_{0}\overline{v_{\theta}^{\prime} s_{1}^{\prime}}$ originating from the non-axisymmetric flows.
It is clearly shown that $\overline{v_{\theta}^{\prime} s_{1}^{\prime}}$ is strongly positive (negative) at high latitudes in the northern (southern) hemisphere, indicating that the high-latitude modes transport heat equatorward.
The amount of equatorward heat transport is more and more enhanced from case1 to case 10.

\section*{Mean states}

\begin{figure}
\begin{center}
\includegraphics[width=0.99\linewidth]{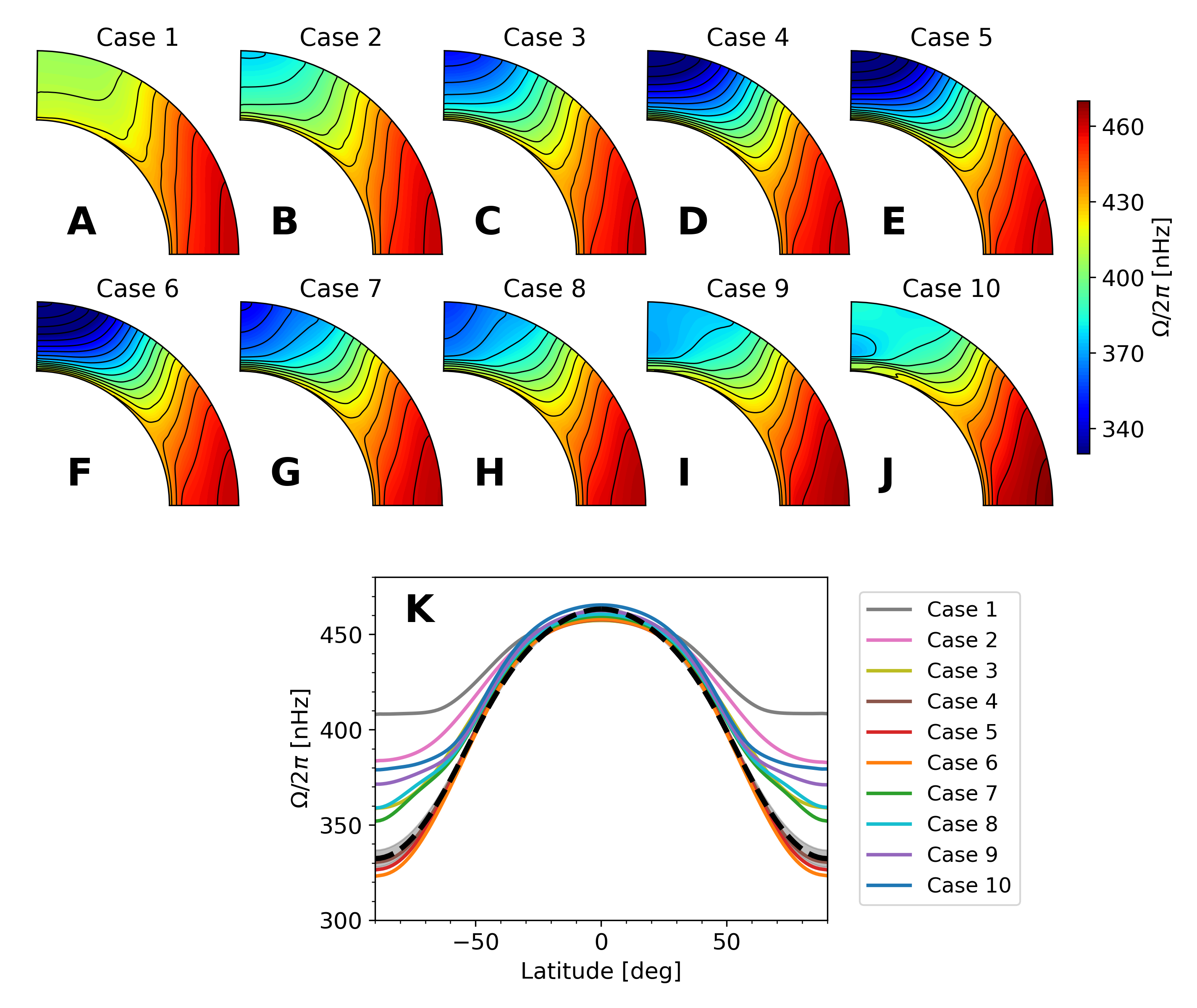}
\caption{ 
{\bf Profiles of the differential rotation ${\Omega}=\Omega_{0}+\overline{v}_{\phi}/(r\sin{\theta})$ in the statistically-stationary states.}
(A--J) Meridional plots of the differential rotation in the northern hemisphere from cases 1--10.
(K) Latitudinal differential rotation in the middle convection zone $r=0.85R_{\odot}$.
The black dashed curve denotes the rate deduced from observations by global helioseismology with GONG data (see Fig.~\ref{fig:1}A).
\label{fig_s:final_DR}
  }
\end{center}
\end{figure}

Figures~\ref{fig_s:final_DR}A--J show the meridional profiles of the differential rotation ${\Omega}=\Omega_{0}+\overline{v}_{\phi}/(r\sin{\theta})$ in the statistically-stationary states.
Cuts in the middle convection zone ($r=0.85R_{\odot}$) are shown in fig.~\ref{fig_s:final_DR}K.
From case 1 to case 6, the latitudinal differential rotation becomes stronger in amplitude and deviates more from the Taylor-Proudman's constraint, i.e., its contour lines are more and more inclined with respect to the rotational axis.
It is seen that our cases 4 to 6 reproduce the latitudinal rotation profile determined by global helioseismology \cite{howe2005}.
In cases 7 to 10, the rotation rates become faster than the observed value at high latitudes by about $30-50$~nHz.

\begin{figure}
\begin{center}
\includegraphics[width=0.99\linewidth]{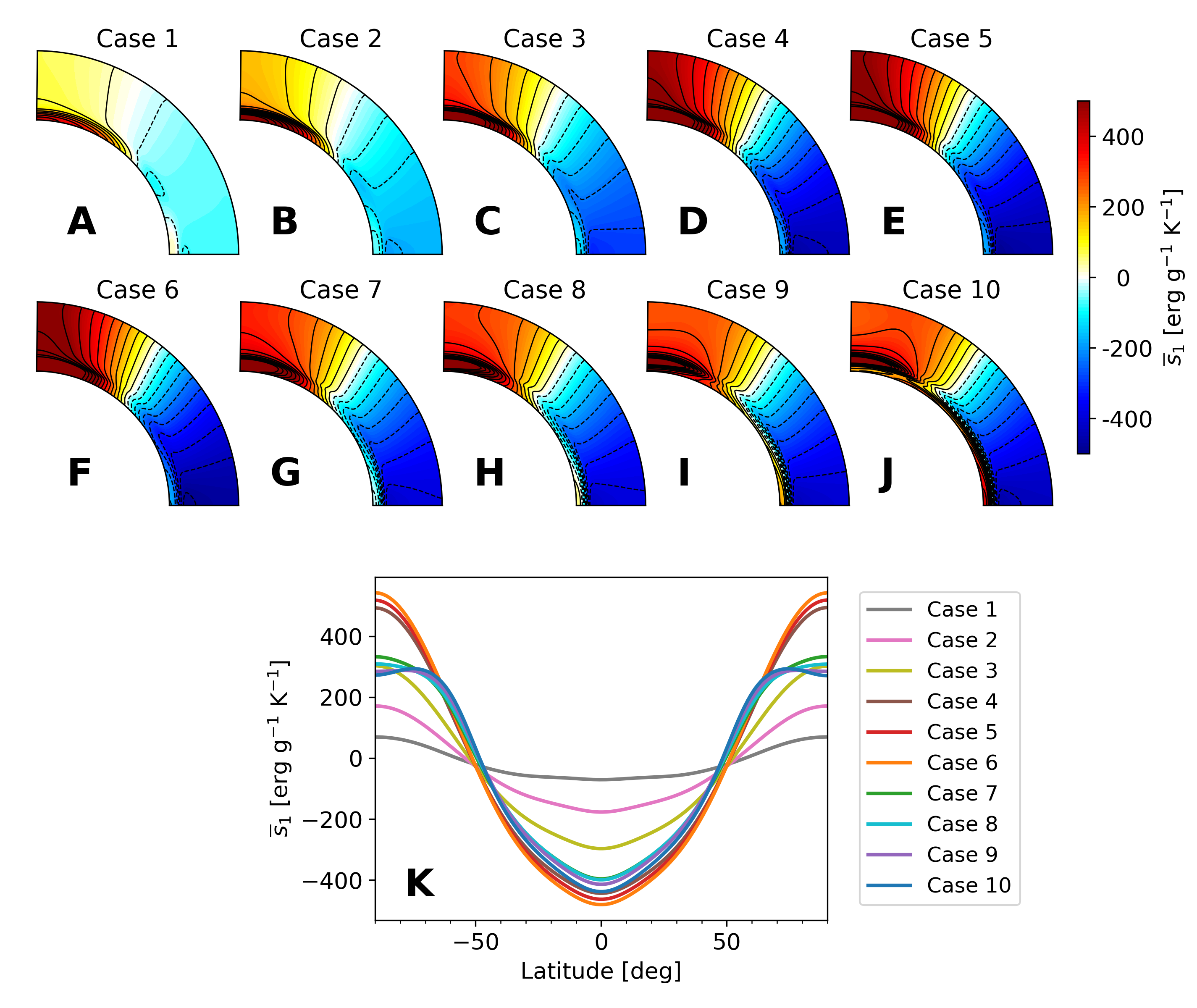}
\caption{ 
{\bf Profiles of the mean entropy perturbation $\overline{s}_{1}$ in the statistically-stationary states.}
(A--J) Meridional plots of $\overline{s}_{1}$ in the northern hemisphere from cases 1--10.
(K) Latitudinal variation of $\overline{s}_{1}$ in the middle convection zone $r=0.85R_{\odot}$.
\label{fig_s:final_twb}
  }
\end{center}
\end{figure}

Figures~\ref{fig_s:final_twb}A--J show the profiles of the mean entropy perturbation $\overline{s}_{1}$ in the statistically-stationary states.
Figure~\ref{fig_s:final_twb}K shows the latitudinal profiles of $\overline{s}_{1}$ in the middle convection zone ($r=0.85R_{\odot}$).
It is seen that, in the middle convection zone, the latitudinal entropy difference (between hotter poles and cooler equator) increases from case 1 to case 6, but drops in cases 7--10 due to the substantial amount of equatorward heat transport by the high-latitude modes.

\begin{figure}
\begin{center}
\includegraphics[width=0.99\linewidth]{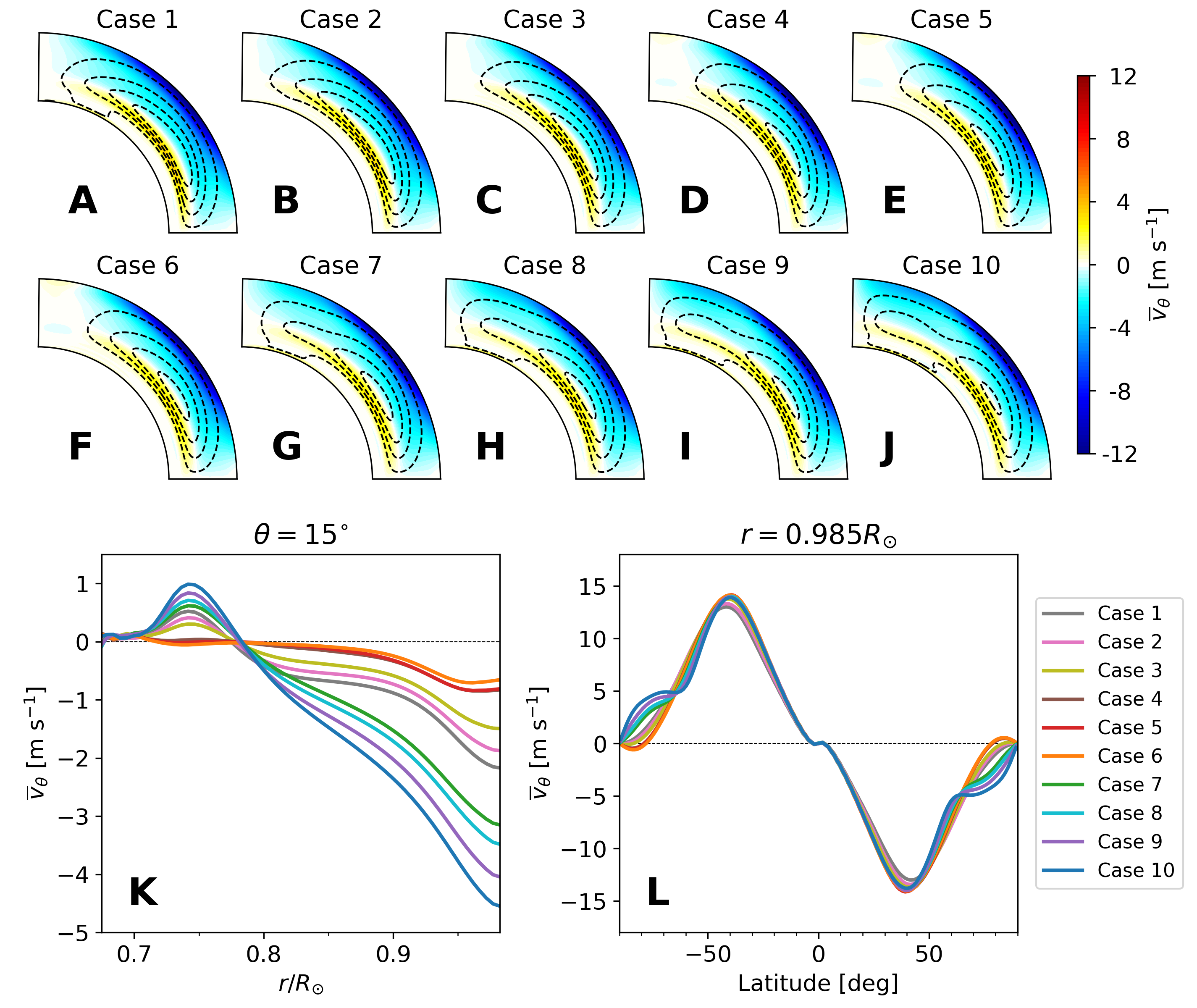}
\caption{ 
{\bf Profiles of the meridional circulation $\overline{\bm{v}}_{\mathrm{m}}=\overline{v}_{r}\bm{e}_{r}+\overline{v}_{\theta}\bm{e}_{\theta}$ in the statistically-stationary states.} 
(A--J) Meridional plots in the northern hemisphere from cases 1--10.
The color maps show the latitudinal velocity $\overline{v}_{\theta}$ whereas the black solid (dashed) lines represent contours of the stream function associated with the clockwise (counter-clockwise) circulation cell.
(K) Latitudinal component of the meridional flow $\overline{v}_{\theta}$ at colatitude $\theta=15^{\circ}$  as a function of radius.
(L) Latitudinal meridional flow $\overline{v}_{\theta}$ at the surface $r=0.985R_{\odot}$ as a function of latitude.
\label{fig_s:final_MC} }
\end{center}
\end{figure}

Figures~\ref{fig_s:final_MC}A--J show the spatial pattern of the meridional circulation $\overline{\bm{v}}_{\mathrm{m}}= \overline{v}_{r}\bm{e}_{r}+\overline{v}_{\theta} \bm{e}_{\theta}$ in the statistically-stationary states.
Blue and red correspond to  poleward and equatorward latitudinal flows $\overline{v}_{\theta}$, respectively, and the dashed lines show the streamlines of counterclockwise meridional circulation cell in the northern hemisphere.
The spatial pattern of the meridional circulation is largely single-cell in each hemisphere in all cases and shows little variations from case 1 to case 10.
This is because the shape of the meridional circulation is dominantly determined by the $\Lambda$ effect (which we fix in all cases) via the so-called gyroscopic pumping \cite{miesch2011,featherstone2015}.
However, some differences exist at high latitudes, as illustrated in figs.~\ref{fig_s:final_MC}K and L.
To understand the cause of these differences, it is convenient to consider the $\phi$-component of the mean vorticity equation which describes the temporal evolution of the meridional circulation
\begin{eqnarray}
    && \frac{\partial \overline{\zeta}_{\phi}}{\partial t} = 
    2r\sin{\theta}\Omega_{0}\frac{\partial {\Omega}_{1}}{\partial z}
    -\frac{g}{r c_{p}}\frac{\partial\overline{s}_{1}}{\partial \theta} 
    + [.\ .\ .], \label{eq_s:vort_phi}
\end{eqnarray}
where $\overline{\zeta}_{\phi}=(\nabla\times\overline{\bm{v}}_{\mathrm{m}})_{\phi}$.
The brackets in the above equation refer to the advective and viscous diffusive terms, which are not important in the following discussion.
In all cases, we find that the meridional circulation largely consists of single counterclockwise circulation cell with $\overline{\zeta}_{\phi}<0 \ (>0)$ in the northern (southern) hemisphere.
This implies that the meridional circulation is primarily driven by the Coriolis force of the differential rotation where $2r\sin{\theta}\Omega_{0} \partial {\Omega}_{1}/\partial z <0 \ (>0)$ in the northern (southern) hemisphere.
From case 1 to case 6, the counterclockwise circulation cell is more and more expelled from the high latitudes and confined in low to middle latitudes.
This is due to the enhancement of the baroclinic torque at high latitudes, $-(g/rc_{\mathrm{p}}) \partial \overline{s}_{1}/\partial\theta$, which counteracts the Coriolis force associated with the differential rotation (see fig.~\ref{fig_s:final_twb}).
In cases 7--10, on the other hand, the counterclockwise cell extends back to higher latitudes (figs.~\ref{fig_s:final_MC}G--J).
This is due to the reduction in the latitudinal entropy difference, which reduces the baroclinic torque.
This promotes the poleward angular momentum transport, leading to the reduction in the differential rotation in cases 7--10 (figs.~\ref{fig_s:final_DR}G--J).

\section*{Comparison with 2D calculations}

In order to assess the impact the high-latitude inertial modes have on the mean state, we  carry out  nonlinear mean-field simulations in a two-dimensional (2D) axisymmetric framework in which the non-axisymmetric modes are not present.
We use the same setup as for the 3D cases with the same $\delta_{0}$ as in the 3D cases 1--10.

The results are reported in Table~\ref{table:1} in parentheses.
In striking contrast to the 3D simulations, we find that in 2D axisymmetric simulations, the latitudinal differential rotation and latitudinal entropy and temperature variations
increase monotonically from case 1 to case 10.

In cases 1--3 where the background baroclinicity is weak, the results from 2D and 3D simulations are very similar.
In cases 4--6 where the high-latitude modes are baroclinically excited to moderate amplitudes,  the differences between the 2D and 3D results
are small
but non-negligible:
For example, in case 5, the latitudinal differential rotation in 3D simulation is smaller than the 2D simulation by about $10$ nHz at high latitudes.
In cases 7--10 where the amplitudes of high-latitude modes are  large, we see very drastic differences between the 2D and 3D simulations.
In case 10, for example, the latitudinal differential rotation and latitudinal temperature difference drop by about $100$ nHz and by $7.4$ K, respectively from the 3D to 2D simulations.
This is due to a substantial amount of equatorward heat transport by the high-latitude modes,  which  changes the the angular momentum transport through the baroclinically-driven meridional flows.

\section*{Angular momentum fluxes} \label{sec:angular}

We now examine the  evolution of the angular momentum fluxes in our nonlinear simulations.
The angular momentum per unit mass is defined as $\mathcal{L}=(r\sin\theta)^2 {\Omega}(r,\theta)$.
We can divide the terms which describe the evolution of $\mathcal{L}$ into different components according to
\begin{eqnarray}
&& \rho_{0} \frac{\partial \mathcal{L}}{\partial t}=-\nabla\cdot (\bm{F}^{\mathrm{MC}}+\bm{F}^{\mathrm{mode}}+\bm{F}^{\Lambda}+\bm{F}^{\mathrm{vis}}),
\end{eqnarray}
where $\bm{F}^{\mathrm{MC}}$, $\bm{F}^{\mathrm{mode}}$, $\bm{F}^{\Lambda}$, and $\bm{F}^{\mathrm{vis}}$ represent the angular momentum fluxes associated with the meridional circulation, the Reynolds stress of the non-axisymmetric flows (largely due to the high-latitude modes), the prescribed $\Lambda$ effect, and the turbulent viscous diffusion, respectively.
These are given by
\begin{eqnarray}
    && \bm{F}^{\mathrm{MC}} = \rho_{0} \overline{\bm{v}}_{\mathrm{m}} \mathcal{L},  \label{eq_s:Fmc} \\
    && \bm{F}^{\mathrm{mode}} = \rho_{0} r\sin{\theta} \overline{\bm{v}_{\mathrm{m}}^{\prime}v_{\phi}^{\prime}},  \label{eq_s:Fmode} \\
    && \bm{F}^{\Lambda} = -\rho_{0} r\sin{\theta} \nu \bm{\Lambda} \Omega_{0},  \label{eq_s:Flambda} \\
    && \bm{F}^{\mathrm{vis}} = -\rho_{0} r^{2}\sin^{2}{\theta} \nu \nabla {\Omega},  \label{eq_s:Fvis}
\end{eqnarray}
where $\bm{v}_{\mathrm{m}}^{\prime}=v_{r}^{\prime} \bm{e}_{r}+v_{\theta}^{\prime} \bm{e}_{\theta}$.
Note that $\bm{\Lambda}=\Lambda_{r\phi} \bm{e}_{r}+\Lambda_{\theta\phi} \bm{e}_{\theta}$ is independent of time.

Figure~\ref{fig_s:instability_case6} shows the radially-averaged latitudinal components of the angular momentum fluxes, $F_{\theta}^{\mathrm{MC}}$, $F_{\theta}^{\mathrm{mode}}$, $F_{\theta}^{\mathrm{\Lambda}}$, $F_{\theta}^{\mathrm{vis}}$, and their sum $F_{\theta}^{\mathrm{tot}}=F_{\theta}^{\mathrm{MC}}+F_{\theta}^{\mathrm{mode}}+F_{\theta}^{\mathrm{\Lambda}}+F_{\theta}^{\mathrm{vis}}$ near the north pole ($\theta=15^{\circ}$) from case 8 as functions of time.
Note that positive (negative) $F_{\theta}$ corresponds to the equatorward (poleward) angular momentum flux in the northern hemisphere.
It is clearly seen that the net poleward angular momentum transport at around $t\approx 17$ yr is caused by the meridional flow ($F_{\theta}^{\mathrm{MC}} <0$).
It is noteworthy that the Reynolds stress $\overline{v_{\theta}^{\prime} v_{\phi}^{\prime}}$ associated with the high-latitude modes is positive (negative) near the north (south) poles, indicating that the modes transport some amount of angular momentum equatorward.
This has been already reported in the solar surface observations \cite{hathaway2013},
and it has sometimes been argued that this Reynolds stress is responsible for accelerating the equator of the Sun \cite{hathaway2013}.
However, our analysis reveals that this equatorward angular momentum transport is over-compensated by the poleward angular momentum transport by the baroclinically-driven meridional flow.
Therefore, the net effect of the baroclinically unstable high-latitude modes on the differential rotation is rather to accelerate the poles relative to the equator.

\section*{Eigenmode extraction}

In order to  characterize the mode properties, the spatial eigenfunctions of the high-latitude modes are extracted from the nonlinear simulations using the singular-value-decomposition (SVD) method \cite{bekki2022b,proxauf2020}.
We use total $16.5$-year-long data with a time cadence of about $5$ days after each simulation reaches a statistically stationary state.
The number of temporal data points is thus $N_{t}=1200$.
We apply Fourier transforms in longitude and time to the longitudinal velocity $v_{\phi}(r,\theta,\phi,t)$ to obtain  
\begin{equation}
\tilde{v}_{\phi}(r,\theta,m,\omega) = \frac{1}{N_{t}N_{\phi}} \sum_{t,\phi} v_{\phi}(r,\theta,\phi,t) e^{\ii(\omega t-m\phi)} .
\end{equation}
The variable $\tilde{v}_{\phi}$ is  a representative variable of the high-latitude modes.
In the above equation $t$, $\phi$ and $\omega$ take discrete values:  $t_{i}=iT/N_{t}$ with $0 \leq i <N_{t}$,  $\phi_{j}=2\pi j/N_{\phi}$ with $0 \leq j <N_{\phi}$, and $\omega_{n}=2\pi n/T$ with $-N_{t}/2 \leq n \leq N_{t}/2$.
In the following analysis, we fix the azimuthal order $m$.
Since we only focus on the modes localized around the poles, we average the spectrum over the middle to high latitudes (HL)
\begin{eqnarray}
    && \tilde{v}_{\phi}^{\mathrm{HL}}(r,\omega) = \frac{4}{\pi}\int_{\theta=0}^{\pi/4} \tilde{v}_{\phi}(r,\theta,\omega) d\theta.
\end{eqnarray}
For each fixed $m$, the high-latitude spectrum is decomposed according to SVD as
\begin{eqnarray}
    && \tilde{v}_{\phi}^{\mathrm{HL}}(r,\omega) = \sum_{k} \sigma_{k}  U_{k}(r) V_{k}^{*} (\omega),
\end{eqnarray}
where $\sigma_{k}$ are the singular values, $U_{k}$ and $V_{k}$ are components of the left and right singular vectors, respectively, and superscript $^{*}$ represents the complex conjugate.
Note that $V_{k}$ is normalized such that $V_{k}^{*}V_{k'}=\delta_{kk'}$.
For eigenmode extraction, we only use the first of the right singular vector, $V_{0}$, whose singular value is dominant over the others.
For each fixed $m$, we use a function $V_{0}(\omega)$ derived from the longitudinal velocity data to calculate the spatial eigenfunction for any variable $q \in \{ v_{r}, \ v_{\theta}, \ v_{\phi}, \ s_{1}, \ p_{1} \}$ as
\begin{eqnarray}
    && q_{\mathrm{mode}}(r,\theta)=\sum_{\omega^{\prime}=-\Omega_{0}}^{0} \tilde{q}(r,\theta,\omega') V_{0}(\omega').
\end{eqnarray}
See section~3.2 in Ref.~\cite{bekki2022b} for more details about the SVD method.

Figure~\ref{fig_s:nonlin_eigfunc}A shows the 1D surface eigenfunctions of longitudinal velocity $v_{\phi}$ of the $m=1$ high-latitude mode extracted from the nonlinear simulation cases 1--10.
Only the northern hemisphere is shown.
In cases 1 and 2, the background baroclinicity is too weak to excite the high-latitude modes.
In cases 3--10, it is seen that the $m=1$ high-latitude mode is self-excited.
The observed amplitude of the $m=1$ mode is consistent with the extracted eigenfucntion in cases 4--6.
On the other hand, the $m=2$ and $m=3$ modes are self-excited only in cases 7--10 and tend to have much smaller amplitudes compared to the $m=1$ mode.
The observed amplitude of the $m=2$ mode is in between the simulation cases 6 and 7, and that of the $m=3$ mode is closest to case~7.

Figure~\ref{fig_s:eigenfunc_case4} shows the structure of the eigenfunctions in the meridional plane for the $m=1$ high-latitude mode from the case 5. 
It is seen that the latitudinal velocity is symmetric across the equator, whereas the radial and longitudinal velocities are north-south antisymmetric across the equator.
This is consistent with the solar observation \cite{gizon2021}.
The associated flow motion is quasi-toroidal, i.e., the radial flow is about $30$ times smaller in amplitudes.
Near the poles, the motions are close to vortical, aligned in the $z$ direction, and in geostrophic balance.
Furthermore, we find that the flow crosses the poles.

\begin{figure}
\begin{center}
\vspace{-1.5cm}
\includegraphics[scale=0.7]{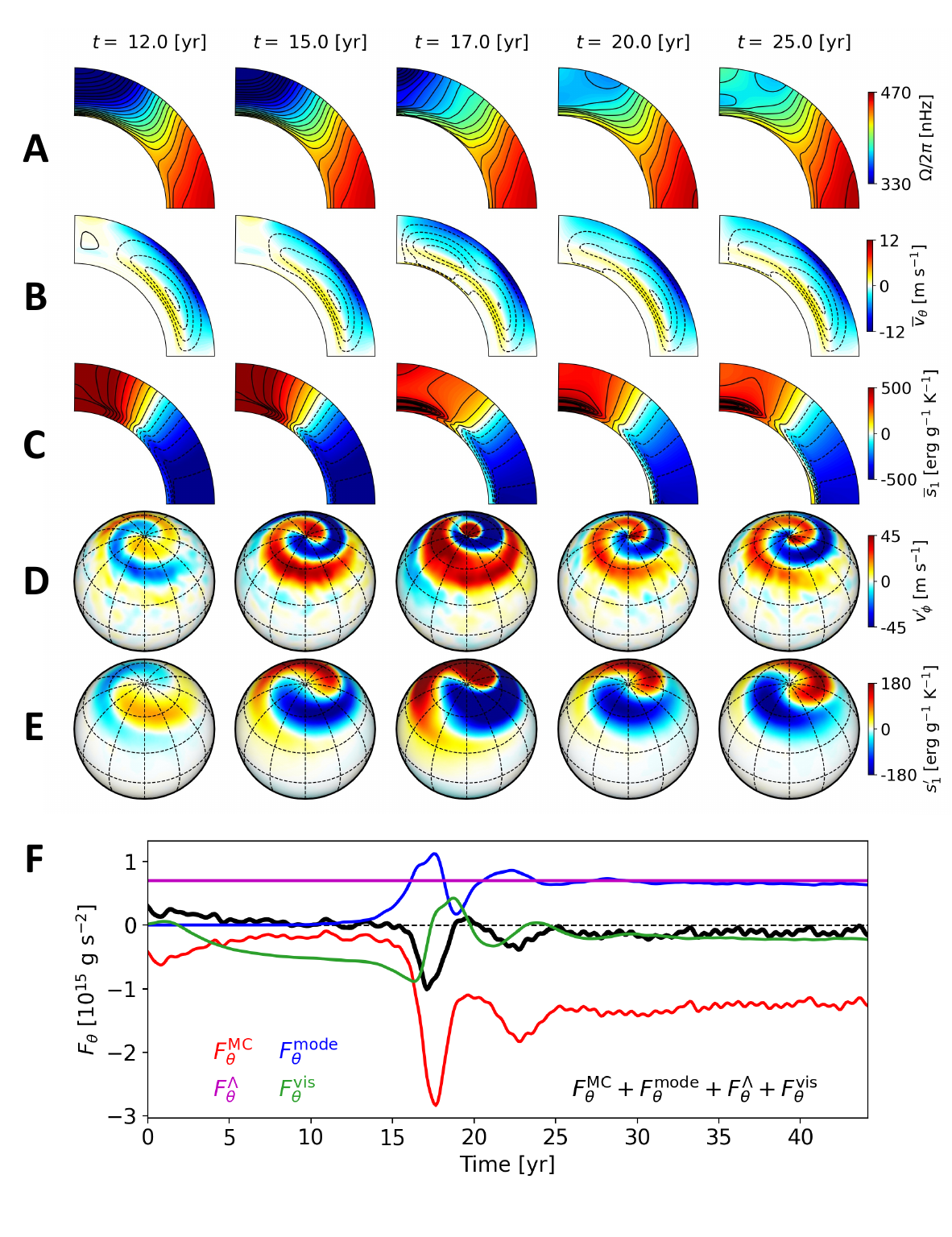}
\caption{
{\bf Nonlinear saturation of the baroclinic instability in the 3D simulation Case~8.}
Snapshots of (A) the differential rotation ${\Omega}$, (B) the latitudinal meridional flow $\overline{v}_{\theta}$, (C) the mean entropy perturbation $\overline{s}_{1}$ are shown in meridional planes.
Snapshots of (D) the longitudinal velocity perturbation $v_{\phi}^{\prime}$, and (E) entropy perturbation $s_{1}^{\prime}$ at the surface are shown.
(F) Temporal evolution of the radially-averaged latitudinal angular momentum fluxes at $\theta=15^{\circ}$ (where positive values indicate equatorward angular momentum transport).
The red, blue, magenta, green, and black curves denote the latitudinal component of the angular momentum fluxes associated with the meridional circulation $F^{\mathrm{MC}}_{\theta}$, the Reynolds stress of the non-axisymmetric modes $F^{\mathrm{mode}}_{\theta}$, the imposed $\Lambda$ effect $F^{\Lambda}_{\theta}$, the turbulent viscous diffusion $F^{\mathrm{vis}}_{\theta}$, and their sum $F^{\mathrm{tot}}_{\theta}$.
Definitions of the various fluxes, $F$,  are given by eqs.~(\ref{eq_s:Fmc})--(\ref{eq_s:Fvis}).}
\label{fig_s:instability_case6}
\end{center}
\end{figure}

\begin{figure}
\begin{center}
\includegraphics[scale=0.36]{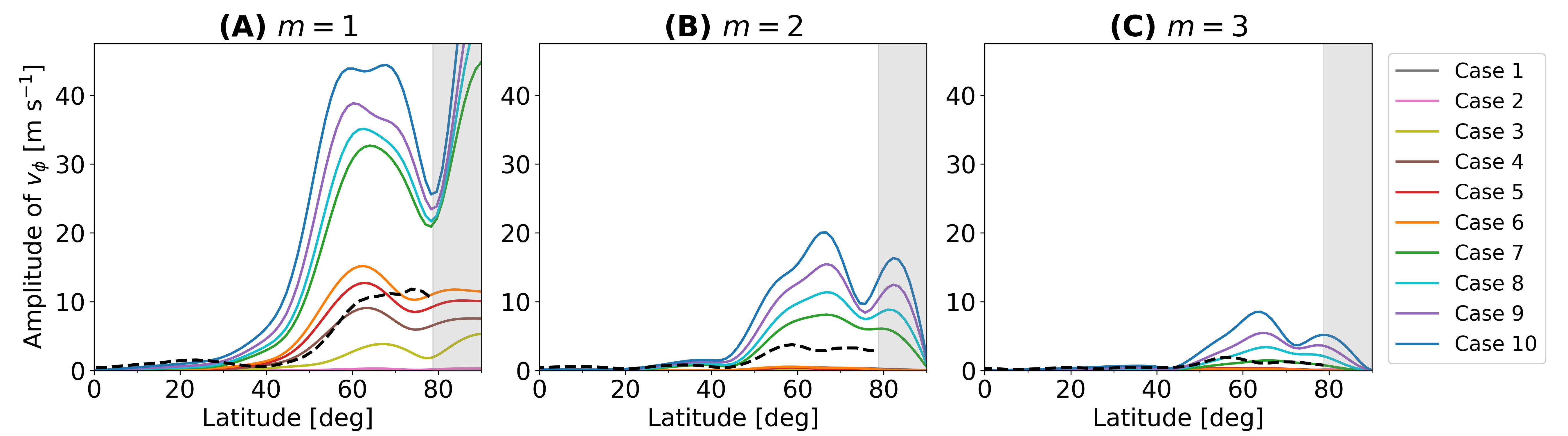}
\caption{ 
{\bf Amplitudes of the longitudinal velocity eigenfunctions of the high-latitude modes extracted from the nonlinear simulations in the northern hemisphere.}
(A--C) Amplitudes of the $m=1$, $2$, and $3$ modes, respectively.
Different colors show the results from the different cases.
The black dashed curve shows the observational results \cite{gizon2021}.
The gray shaded area indicates the range of latitudes where the observations are not available.
\label{fig_s:nonlin_eigfunc}
  }
\end{center}
\end{figure}

\begin{figure}
\begin{center}
\includegraphics[scale=0.2]{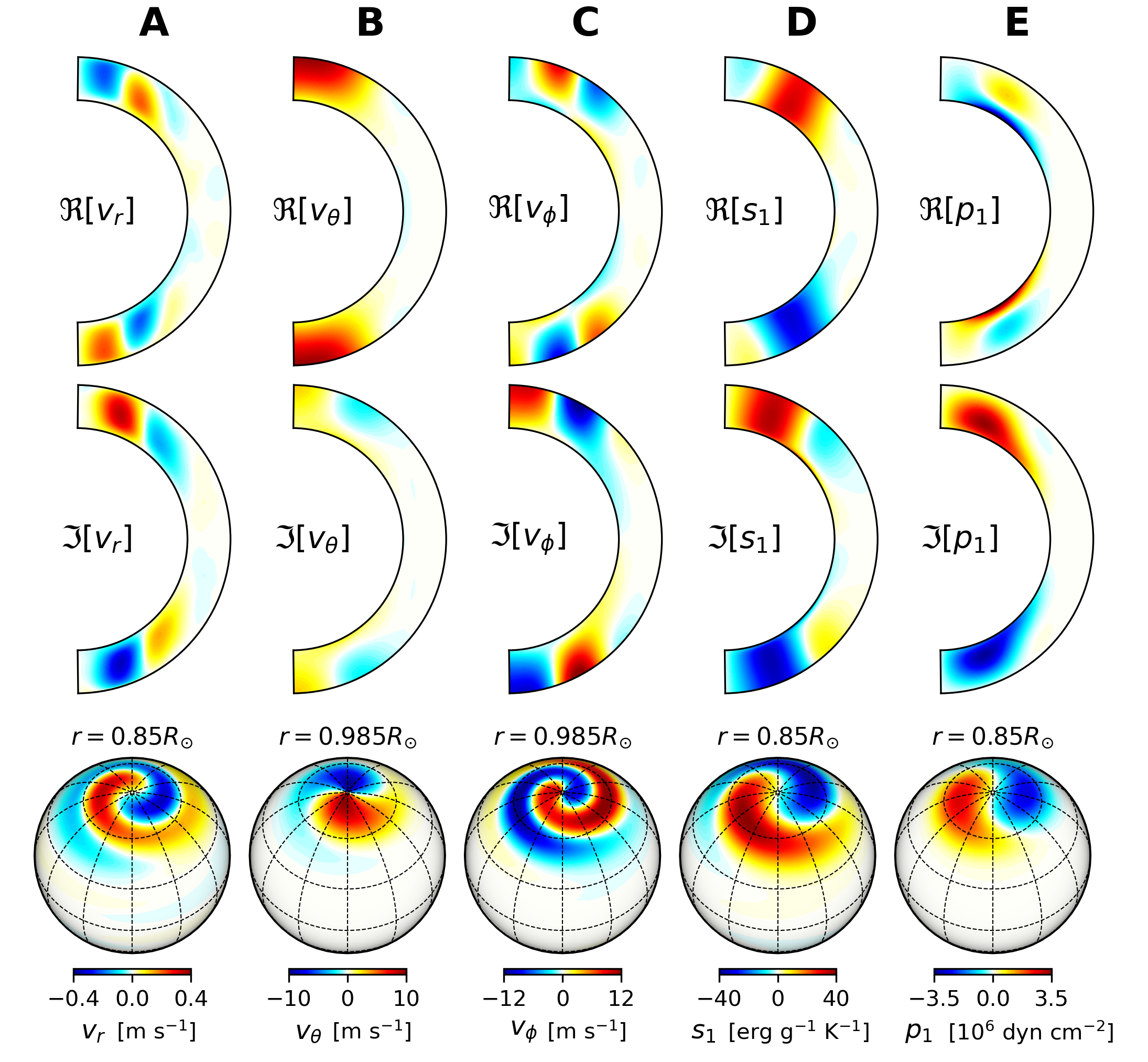}
\caption{
{\bf Extracted eigenfunctions of the $m=1$ high-latitude mode from the nonlinear simulation case 5.} The figures show the real and imaginary parts of the eigenfunctions for (A) the radial velocity $v_{r}$, (B) the latitudinal velocity $v_{\theta}$, (C) the longitudinal velocity $v_{\phi}$, (D) the entropy perturbation $s_{1}$,  and (E) the pressure perturbation $p_{1}$.
The bottom row shows cuts through the eigenfunctions $\Re[q_{\rm mode}(r,\theta) e^{\ii \phi}]$ at fixed radii $r$, where $q_{\rm mode}$ refers to either $v_r$, $v_\theta$, $v_\phi$, $s_1$, or $p_1$.
The eigenfunctions are shown at the surface, $r=0.985R_{\odot}$, for $v_{\theta}$ and $v_{\phi}$  (where they reach their maximum values) and in the middle of the convection zone, $r=0.85R_{\odot}$, for $v_{r}$, $s_{1}$, and $p_{1}$ (where they approximately reach their maximum values).
\label{fig_s:eigenfunc_case4}
 }
\end{center}
\end{figure}

As already discussed, the high-latitude modes play important roles in redistributing the angular momentum and heat in the convection zone.
Figures~\ref{fig_s:RSFehk_case4}A and B show meridional profiles of the horizontal Reynolds stress $\rho_{0}\overline{v_{\theta}v_{\phi}}$ and the latitudinal component of the thermal energy flux $\rho_{0}c_{p} \overline{v_{\theta} T_{1}}$ associated with the $m=1$ high-latitude mode extracted from the simulation case 5.
The Reynolds stress is positive (negative) in the northern (southern) hemisphere in the bulk of the convection zone, implying the the high-latitude mode transports the angular momentum equatorward.
This is consistent with the observations made by Refs.~\cite{hathaway2013}.
Similarly, the latitudinal thermal energy flux is directed equatorward in both hemispheres, indicating that the mode also transports the heat equatorward.
We find that this equatorward heat transport by the high-latitude modes affects the meridional flow and thus its angular momentum transport. This more than compensates the well-known equatorward angular momentum transport (fig.~\ref{fig_s:instability_case6}).

\section*{Dependence on $\Lambda$ effect}

In this study, we have fixed the $\Lambda$ effect and only varied the subadiabaticity at the base of the convection zone to control the baroclinicity in the system.
To demonstrate the robustness of the results, we present in this section a sets of additional mean-field simulations with different $\Lambda$-effect parameters.
The profiles of the large-scale mean flows sensitively depend on the direction $\lambda$  of the angular momentum flux of the $\Lambda$ effect \cite{rempel2005,bekki2017a}.
Under the observational constraints of the solar-like differential rotation (faster equator and slower poles) and the single-cell meridional circulation in each hemisphere (poleward flow at the surface and equatorward flow at the base), the direction $\lambda$ is required to be  equatorward.
Within this limitation, the change in $\lambda$ has a minor impact of changing the amplitudes of the large-scale mean flows \cite{rempel2005}.
Thus, in this section, we fix the direction $\lambda$ and only vary the amplitude of the $\Lambda$ effect, $\Lambda_0$.

With the larger $\Lambda_{0}$, the angular momentum is more strongly transported equatorward, leading to a stronger latitudinal differential rotation.
Figure~\ref{fig_s:Omeq_lambda0} shows the equatorial rotation rate at the top surface ($r=0.985R_{\odot}$)
as a function of $\Lambda_{0}$.
Here, the subadiabaticity at the base of the convection zone is fixed to the value from the reference case 5, $\delta_{0}=-2.2\times 10^{-5}$.
Given that the observed equatorial rotation rate of the Sun is $\approx 460$ nHz at $r=0.985R_{\odot}$, we can safely exclude too small ($< 0.4$) or too large ($> 1.2$) values of $\Lambda_{0}$.

Here, we have additionally carried out sets of 3D mean-field simulations with varying $\delta_{0}$ for two additional values of $\Lambda_{0}$ ($0.75$ and $0.95$ in addition to the reference value $\Lambda_{0}=0.85$).
The results are summarized in table~\ref{table_s:lambda} and shown in fig.~\ref{fig_s:fig4_lambda}.
When $\Lambda_{0}$ is larger (smaller), the latitudinal temperature difference $\Delta_{\theta}\overline{T}_{1}$ becomes larger (smaller) for the same subadiabaticity $\delta_{0}$ at the base of the convection zone.
This is because the meridional flow, which is largely determined by gyroscopic pumping \cite{featherstone2015}, becomes stronger (weaker) for larger (smaller) $\Lambda_{0}$, leading to a more (less) efficient generation of the entropy perturbation at the base of the convection zone.
Hence, when $\Lambda_{0}$ is small, the base of the convection zone needs to be more subadiabatic in order to achieve the same latitudinal temperature difference $\Delta_{\theta}\overline{T}_{1}$.
Nonetheless, it is seen that, regardless of $\Lambda_{0}$, all the simulations follow a very similar trend:
As the base of the convection zone becomes more subadiabatic, the latitudinal temperature difference $\Delta_{\theta}\overline{T}_{1}$ increases and the amplitudes of the high-latitude modes increase.
When the mode amplitudes become sufficiently large, $\Delta_{\theta}\overline{T}_{1}$ is substantially reduced due to the nonlinear feedback from the modes.
This is accompanied by a reduction in the latitudinal differential rotation $\Delta_{\theta}{\Omega}$.
Under the observational constraints on $\Delta_{\theta}{\Omega}$ and the high-latitude mode amplitudes, $\Delta_{\theta}\overline{T}_{1}$ is limited to be less than $7$~K.
Therefore, our general conclusion is robust to the amplitude of the  $\Lambda$ effect.

\section*{Potential implications for the dynamo}

Although the magnetic field has not been included in our numerical simulations,
we offer here some additional comments on the potential interactions between the high-latitude inertial modes and the solar dynamo (beyond the fact that the modes affect the differential rotation).

\begin{figure}
\begin{center}
\includegraphics[scale=0.6]{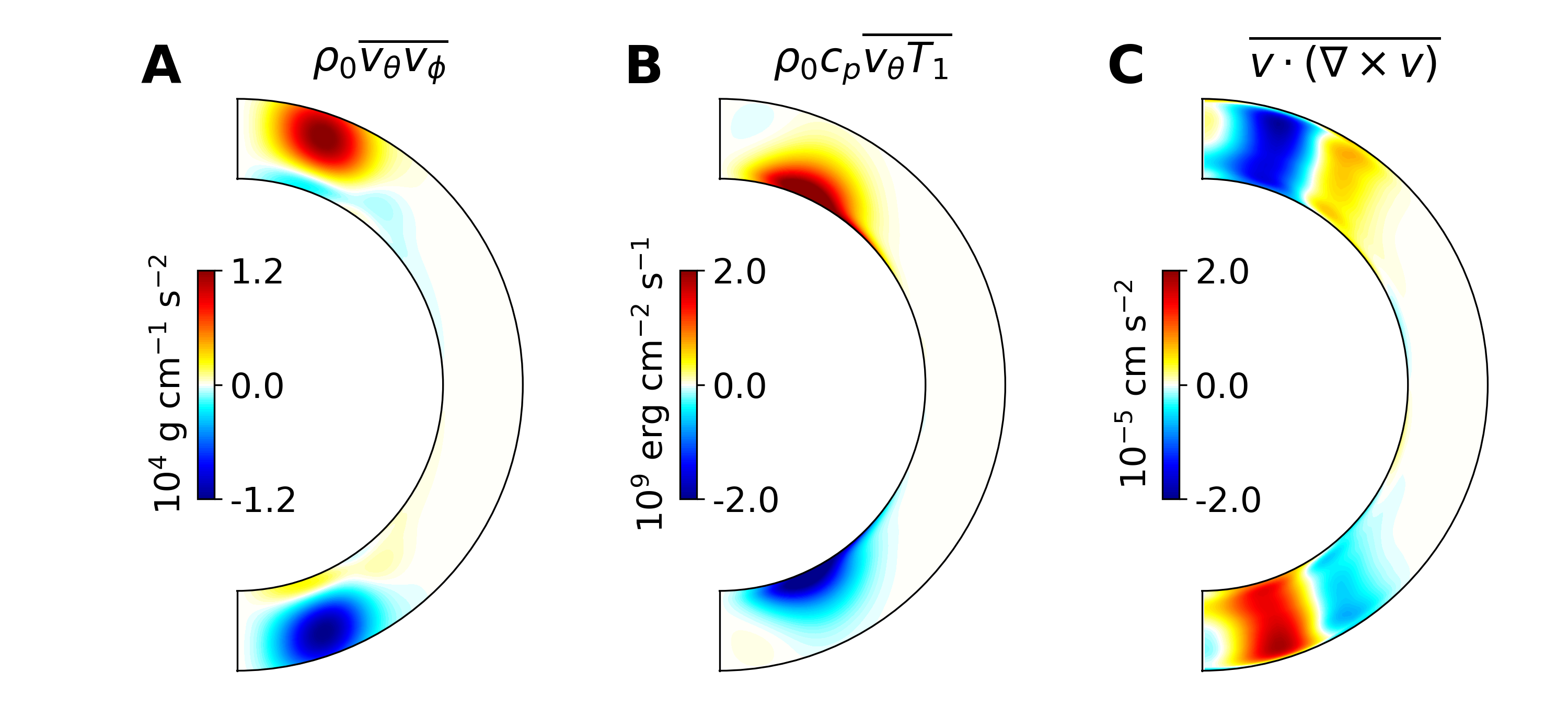}
\caption{
{\bf Properties of the $m=1$ high-latitude mode extracted from our nonlinear simulation Case 5.}
(A) Horizontal Reynolds stress $\rho_{0} \overline{ v_{\theta}v_{\phi}}$.
(B) Latitudinal enthalpy flux $\rho_{0}c_{p} \overline{ v_{\theta}T_{1}}$.
(C) Kinetic helicity $h_{\mathrm{k}}=\overline{ \bm{v}\cdot (\nabla\times\bm{v}) }$.
\label{fig_s:RSFehk_case4}
  }
\end{center}
\end{figure}

\begin{figure}
\begin{center}
\includegraphics[scale=0.7]{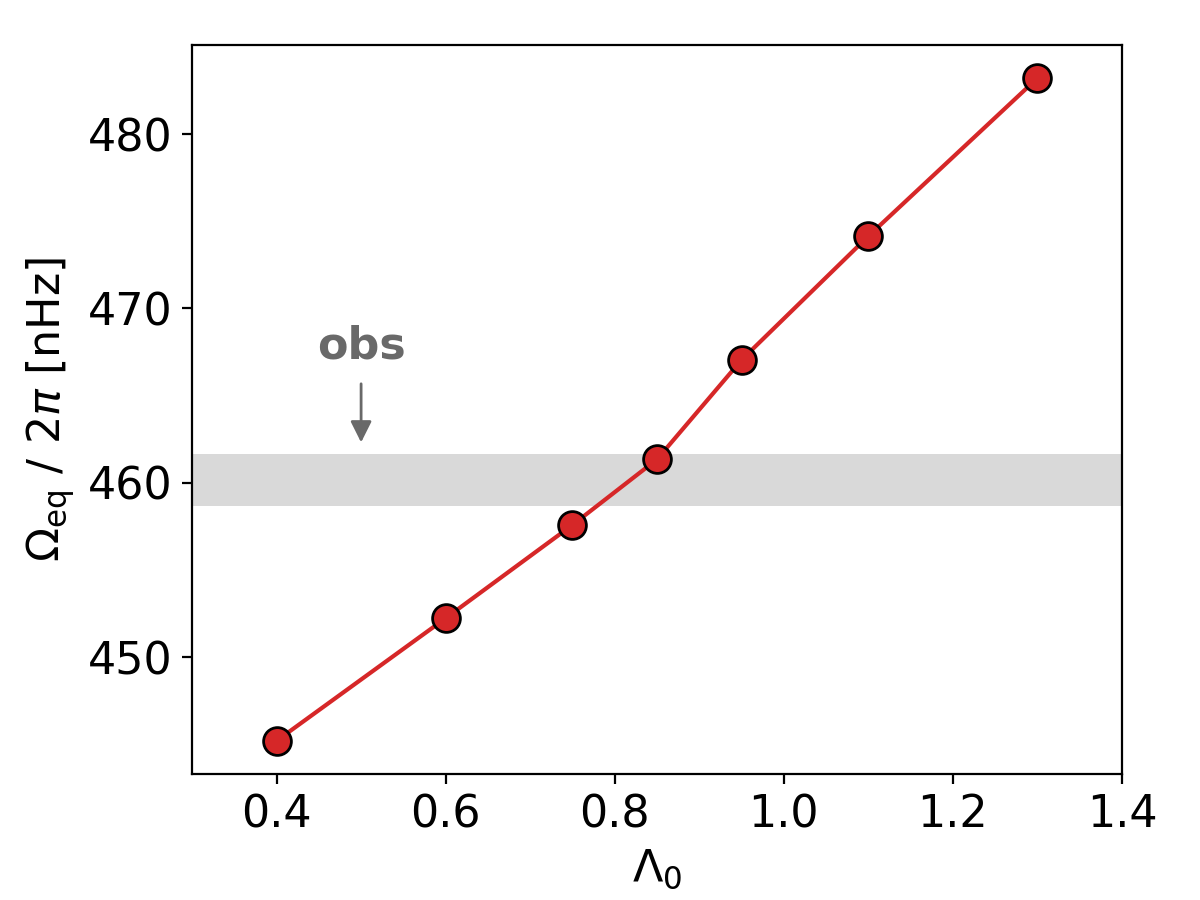}
\caption{
{\bf Results from the parameter survey on amplitudes of the $\Lambda$ effect, $\Lambda_{0}$.}
The equatorial rotation rates at the surface ${\Omega}_{\mathrm{eq}}={\Omega}(r=0.985R_{\odot},\theta=\pi/2)$ in statistically-stationary states are plotted for different values of $\Lambda_{0}$.
The shaded region shows the observed  ${\Omega}_{\rm eq}$.
\label{fig_s:Omeq_lambda0}
  }
\end{center}
\end{figure}

\begin{figure}
\begin{center}
\includegraphics[scale=0.62]{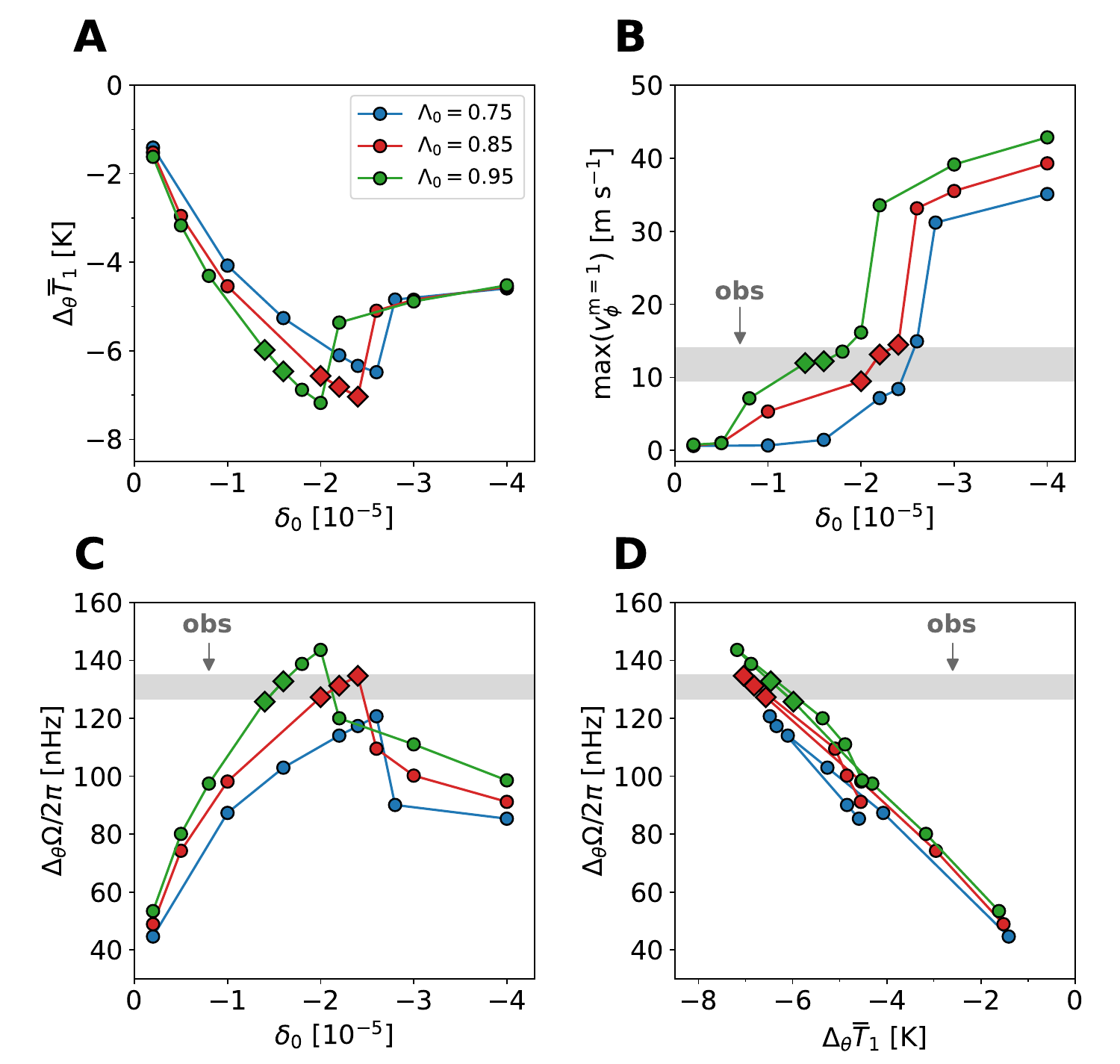}
\caption{
{\bf Comparison of the simulation results for different values of the $\Lambda$-effect amplitude.}
The blue, red, and green markers show results from the cases $\Lambda_{0}=0.75$, $0.85$ (the same as in Fig.~\ref{fig:4}), and $0.95$ respectively.
({\bf A})  $\Delta_{\theta}\overline{T}_{1}$ at $r=0.85R_{\odot}$ as a function of $\delta_{0}$.
({\bf B}) Longitudinal velocity amplitude of the $m=1$ mode at the surface.
({\bf C}) Latitudinal differential rotation $\Delta_{\theta}{\Omega}$ at $r=0.85 R_{\odot}$.
({\bf D}) Relationship between $\Delta_{\theta}\overline{T}_{1}$ and $\Delta_{\theta}{\Omega}$.
The gray shaded areas show the observed values.
}
\label{fig_s:fig4_lambda}
\end{center}
\end{figure}

We find that a substantial negative (positive) kinetic helicity $\overline{\bm{v}^{\prime}\cdot(\nabla\times\bm{v}^{\prime})}$ is associated with the high-latitude inertial modes near the northern (southern) pole (fig.~\ref{fig_s:RSFehk_case4}C). This might contribute to the large-scale dynamo processes at high latitudes.
Furthermore, the Sun's magnetic field potentially affects the baroclinicity in several ways.
On the one hand, the small-scale magnetic fields are expected to enhance the anisotropic convective heat transport and to increase the latitudinal temperature gradient in the convection zone \cite{hotta2018}, which can further enhance the baroclinic excitation of the high-latitude modes.
On the other hand, some studies imply that the large-scale toroidal field has a general tendency to suppress the baroclinic instability \cite{gilman2015,gilman2017}.
Further simulations are required to assess the dynamical role of the baroclinically unstable inertial modes on the Sun's differential rotation in the presence of magnetic fields. 

\clearpage
\newpage







\renewcommand{\arraystretch}{1.25}
\begin{table}[h!]
 \begin{center} 
\caption{
{\bf Results from supplementary simulations with $\Lambda$-effect amplitudes that are smaller ($\Lambda_{0}=0.75$) and larger ($\Lambda_{0}=0.95$) than the value from Table~1 ($\Lambda_{0}=0.85$).}
}
\small
\begin{tabular}{ccccccccccc} 
\hline
\hline
\multirow{2}{*}{$\Lambda_{0}$}  & \multirow{2}{*}{$\delta_{0}$} & $\Delta_{\theta}{\Omega}/2\pi$ 
    & $\Delta_{\theta} \overline{s}_{1}$
    & $\Delta_{\theta} \overline{T}_{1}$ & \multicolumn{4}{c}{max($v_{\phi}$)  [m s$^{-1}$]} \\
\cline{6-11}
&& [nHz] & [erg g$^{-1}$ K$^{-1}$] & [K] & $m=1$ & $m=2$ & $m=3$   \\
 \hline
$0.75$ & $-2.0\times 10^{-6}$ & $44.6$ & $-132$ & $-1.4$ &  $0.6$ & $0.4$ & $0.4$   \\
$0.75$ & $-1.0\times 10^{-5}$ & $87.3$ & $-539$ & $-4.1$ &  $0.7$ & $0.5$ & $0.4$   \\
$0.75$ & $-1.6\times 10^{-5}$ & $102.9$ & $-736$ & $-5.3$ &  $1.5$ & $0.6$ & $0.5$   \\
$0.75$ & $-2.2\times 10^{-5}$ & $114.0$ & $-878$ & $-6.1$ &  $7.2$ & $0.8$ & $0.6$   \\
$0.75$ & $-2.4\times 10^{-5}$ & $117.3$ & $-919$ & $-6.3$ &  $8.4$ & $0.8$ & $0.6$   \\
$0.75$ & $-2.6\times 10^{-5}$ & $120.7$ & $-947$ & $-6.5$ &  $14.9$ & $1.1$ & $0.6$   \\
$0.75$ & $-2.8\times 10^{-5}$ & $90.1$ & $-689$ & $-4.8$ &  $31.2$ & $7.9$ & $2.1$   \\
$0.75$ & $-4.0\times 10^{-5}$ & $85.3$ & $-689$ & $-4.6$ &  $35.1$ & $9.9$ & $3.0$   \\
\hdashline[1pt/1pt]
$0.95$ & $-2.0\times 10^{-6}$ & $53.4$ & $-146$ & $-1.6$ &  $0.8$ & $0.5$ & $0.6$   \\
$0.95$ & $-5.0\times 10^{-6}$ & $80.1$ & $-375$ & $-3.2$ &  $1.0$ & $1.0$ & $1.2$   \\
$0.95$ & $-8.0\times 10^{-6}$ & $97.5$ & $-549$ & $-4.3$ &  $7.1$ & $1.2$ & $0.6$   \\
$0.95$ & $-1.4\times 10^{-5}$ & $125.8$ & $-819$ & $-5.9$ &  $11.9$ & $1.0$ & $0.7$   \\
$0.95$ & $-1.6\times 10^{-5}$ & $132.8$ & $-899$ & $-6.5$ &  $12.2$ & $1.1$ & $0.7$   \\
$0.95$ & $-1.8\times 10^{-5}$ & $138.8$ & $-971$ & $-6.9$ &  $13.5$ & $1.1$ & $0.7$   \\
$0.95$ & $-2.0\times 10^{-5}$ & $143.6$ & $-1025$ & $-7.2$ &  $16.1$ & $1.2$ & $0.8$   \\
$0.95$ & $-2.2\times 10^{-5}$ & $120.0$ & $-763$ & $-5.4$ &  $33.6$ & $7.9$ & $2.0$   \\
$0.95$ & $-3.0\times 10^{-5}$ & $111.0$ & $-724$ & $-4.9$ &  $39.1$ & $12.4$ & $4.1$   \\
$0.95$ & $-4.0\times 10^{-5}$ & $98.6$ & $-712$ & $-4.5$ &  $42.8$ & $15.2$ & $6.2$   \\
\hline
\end{tabular}
\label{table_s:lambda}
\end{center}
\end{table}
\renewcommand{\arraystretch}{1.0}



\end{document}